# Beyond the Coast: an Empirical Assessment of the Kaldor-Verdoorn Law in Chinese Provinces

Maria Cristina Barbieri Góes

Department of Economics and Business Studies (DiSEI), University of Eastern Piedmont, Novara, Italy

Saverio Barabuffi

Institute of Management and L'EMbeDS Department, Sant'Anna School of Advanced Studies, Pisa, Italy

**Abstract**

This paper investigates spatial productivity convergence across Chinese provinces during structural transformation, examining how a shift in autonomous demand composition reshaped regional productivity dynamics. Using Panel Structural Vector Autoregressive modelling and province-level data from 2001 to 2021, we analyse eastern, central, and western regions across two sub-periods through the augmented Kaldor-Verdoorn law. A post-2011 spatial reversal emerges: inland regions exhibit stronger Verdoorn coefficients while fiscal multipliers converge across coastal and inland areas, validating China's redistributive fiscal policies in narrowing regional productivity gaps. The Kaldor-Verdoorn coefficient remains robust throughout, while export effects on productivity collapse and government expenditure effects strengthen.

## 1 Introduction

The recent stagnation in labor productivity growth in advanced economies has prompted renewed interest in understanding the mechanisms driving long-term productivity dynamics (Antenucci et al., 2020; Goldin et al., 2024; Fernald et al., 2025). While growth regimes provide a useful framework for understanding the drivers of output growth (Baccaro and Pontusson, 2016; Hassel and Palier, 2021), they do not theorize any dynamic relationship between aggregate demand and productivity. To address this limitation, we complement the growth regime framework with a demand-led perspective emphasizing the endogenous relationship between demand, output expansion, and productivity improvements (Fazzari et al., 2020; Deleidi et al., 2021, 2023; Fazzari and González, 2025). This perspective, articulated through the Kaldor-Verdoorn law, posits that productivity growth emerges from output growth through economies of scale, division of labour, and learning-by-doing (Kaldor, 1966; Michl, 1985), with supply endogenously accommodating demand dynamics through hysteresis effects (Fazzari et al., 2020; Fazzari and González, 2025).

China's economic transformation since WTO accession in 2001 provides a compelling empirical context for studying how different growth drivers translate into productivity outcomes via the Kaldor-Verdoorn mechanism. The initial export-led model concentrated economic activity in eastern coastal regions, creating substantial productivity gaps between coastal and inland areas (Zhao and Tong, 2000). A subsequent reversal began in the mid-2000s as inland regions experienced accelerated industrial development and productivity convergence (Lemoine et al., 2015), reflecting major policy initiatives promoting balanced regional development (Li et al., 2025). Recent work has documented heterogeneous effects of fiscal policy across Chinese provinces (He and Yu, 2025), underscoring the need for frameworks that identify the specific channels through which government expenditure translates into productivity gains, a gap the Kaldor-Verdoorn approach directly addresses.

Building on Di Carlo et al. (2024) and Regan and Blyth (2025), we analyse subnational growth regimes in a large transitional economy, linking demand-side components to productivity dynamics through the Kaldor-Verdoorn law. Using P-SVAR modelling on Chinese provincial data from 2001 to 2021, we examine three interrelated dimensions: first, how structural shocks drive output growth and labour productivity through the Kaldor-Verdoorn mechanism; second, how China's transition from export-led to domestically-oriented development altered productivity dynamics; and third, how spatial convergence patterns across eastern, central, and western regions reflect the effectiveness of redistributive fiscal policies. The remainder of this paper is organised as follows. Section 2 reviews the theoretical and empirical foundations. Section 3 presents the data and methodology. Section 4 reports the empirical findings. Section 5 concludes, summarazing our findings.

# 2 Literature Review

This section reviews the theoretical and empirical foundations that supports our analysis of regional productivity dynamics in China's transition economy. We organize the literature around two interconnected themes that directly inform our research questions and empirical strategy. First, we examine growth regimes alongside the Kaldor-Verdoorn law and its recent extensions, which provide the theoretical framework for understanding how demand, particularly through its autonomous components (i.e. exports and government expenditure), drives endogenous productivity growth. Second, we review empirical studies on China's regional development patterns, focusing on how export-led growth and government policies have shaped spatial productivity differentials and regional convergence trajectories since WTO accession in 2001.

## 2.1 Bridging Growth Regimes and Kaldorian Dynamics

As proposed by Baccaro and Pontusson (2016), growth regimes identify the dominant components of aggregate demand (exports, government expenditure, or private consumption) that sustain output growth in a given national or subnational economy, treating labour productivity as endogenous to that demand configuration. Yet the growth regimes framework operates primarily at the level of steady-state accounting: it identifies which demand components prevail, but offers no dynamic theory of how shifts in those components translate into productivity outcomes over time. This is a significant limitation, because the policy implications of a transition from export-led to fiscally-driven growth depend critically on whether, and through which mechanisms, different demand components generate productivity gains. The Kaldor-Verdoorn Law directly addresses this gap by formalising the positive, endogenous relationship between output expansion and productivity growth (Verdoorn, 1949; Kaldor, 1966), identifying productivity improvements as endogenously generated by demand-driven output expansion through economies of scale, division of labour, and learning-by-doing (Kaldor, 1966; Michl, 1985).

Recent contributions have refined the Kaldor-Verdoorn framework by decomposing aggregate demand into autonomous and induced components, identifying exports and government expenditure as the primary drivers of accumulation (Fazzari et al., 2020; Fazzari and González, 2025). Strong autonomous demand reduces economic slack and stimulates productivity through hysteresis effects, where supply endogenously accommodates demand dynamics, directly challenging the neoclassical supply-side constraint view. This suggests a dynamic complementarity between growth regimes and Kaldorian productivity theory: the specific configuration of a growth regime determines the channels through which productivity improvements are generated, while the Kaldor-Verdoorn mechanism quantifies their magnitude. A transition between growth regimes should therefore be detectable in the shifting relative effectiveness of exports and government expenditure as drivers of productivity, precisely what our empirical framework is designed to identify.

Cross-country evidence from Antenucci et al. (2020) confirms robust Verdoorn coefficients (0.4 to 0.6). At the regional level, the law has been validated across European regions (Pons-Novell and Viladecans-Marsal, 1999; Barbieri Góes et al., 2026), with subnational studies on Italian regions revealing heterogeneous productivity responses to demand shocks (Deleidi et al., 2021). Crucially, Deleidi et al. (2023) demonstrate that the productivity-growth link is regime-dependent: export shocks catalyse stronger responses in developed economies, whereas government expenditure acts as the primary engine in developing contexts.

Applying this combined framework at the subnational level is both theoretically motivated and empirically necessary. The socio-economic heterogeneity of large countries necessitates shifting the unit of analysis from national to subnational entities (Di Carlo et al., 2024; Regan and Blyth, 2025), while Barbieri Góes et al. (2026) demonstrate that productivity dynamics exhibit significant non-linearities obscured in aggregate national estimates. If growth regimes vary across regions and the Kaldor-Verdoorn mechanism is sensitive to regime composition, national-level estimates will systematically misrepresent individual regional dynamics. Yet both bodies of literature have been applied exclusively to advanced economies and European regions, their application to large emerging economies undergoing simultaneous structural transformation and regional divergence is entirely absent, notwithstanding earlier static applications of the Kaldor-Verdoorn law to China (Hansen and Zhang, 1996; Guo et al., 2013).

China provides an unusually demanding test of this combined framework, having undergone a documented transition between growth regimes within our sample period (from export-led coastal industrialisation toward domestically-oriented, fiscally-anchored development) allowing us to observe how the Kaldor-Verdoorn mechanism responds to a fundamental shift in autonomous demand composition. Earlier applications (Hansen and Zhang, 1996; Guo et al., 2013), while establishing that industrial growth and spatial spillovers drive regional income, treat the law as a static aggregate technical constant rather than a dynamic outcome of a specific demand-side regime, and cannot explain whether productivity gains are driven by global export markets or government-led fiscal expansion — the central question this paper addresses.

## 2.2 China's Regional Development and Spatial Productivity Patterns

Since China's accession to the WTO in 2001, the country began a process of catching-up through an export-led regime characterised by rapid growth of labour productivity in the manufacturing sector (Yu et al., 2015), reaching annual rates of 2.85% for gross output and 7.96% for value-added between 1998-2006 (Brandt et al., 2012). The international integration of the Chinese economy fueled productivity growth particularly in regions which pursued export-driven growth strategies (Erten and Leight, 2021) and leveraged inward Foreign Direct Investments (FDI) which facilitated learning and technology spillovers (Brandt et al., 2017). This initial phase of export-led development concentrated economic activity and productivity gains in eastern coastal provinces, establishing the spatial patterns that subsequent policies sought to rebalance (Zhao and Tong, 2000).

### 2.2.1 Export-Led Growth, Government Expenditure and Regional Convergence

China's trade and investment liberalization initially led to substitution of domestic production for imported materials by individual processing exporters (Kee and Tang, 2016) which crowd out domestic investment (Chen et al., 2017) and contributed to regional inequality (Duan et al., 2023). Moreover, FDI flows retarded labour productivity growth and distorted the industrial structure of Chinese regions (Lo et al., 2016), even though Chinese employment growth was mainly driven by domestic demand rather than exports (Los et al., 2015). These findings suggest that the export-productivity nexus, while positive in aggregate, generated uneven spatial benefits.

Although Chinese regional economic growth divergence depended on the different degrees of economic openness to international trade and varying commitments to market reforms (Lin et al., 2013), a fundamental shift began after the 2007-2008 financial crisis when domestic inputs were increasingly produced by domestic enterprises with very limited foreign capital (Duan et al., 2021), and the export slowdown in China during the mid-2010s led local party secretaries to increase expenditures on public security and social spending (Tombe and Zhu, 2019).

In response to the spatial imbalances, Chinese central government policies increasingly emphasized redistributive fiscal transfers and targeted regional development programs. On the one hand, China's central government spending in the provinces boosted regional economic growth and narrowed regional economic disparities through infrastructure investment and preferential policies for inland regions (Feng et al., 2024). By expanding fiscal resources available to inland provincial governments, these redistributive transfers enabled capacity for sustained economic governance improvements (Wilson, 2016). On the other hand, the effectiveness of these policies remains contested. Liu et al. (2022) argue that fiscal transfers may have reduced output growth in wealthy provinces by decreasing local government incentives to promote development, while Wu and Wang (2013) find a negative relationship between transfer dependency and expenditure decentralization in China, suggesting that intermediate governments might use central transfers to enhance the growth of public employment for their own interests rather than productive investment.

Despite these potential inefficiencies, the literature documents multiple channels through which Chinese regions achieved productivity convergence. Long-term growth convergence relied mainly on accelerating capital accumulation rates as a growth strategy in "catching-up" regions (Andersson et al., 2013), yet Zhang et al. (2019) demonstrate that growth convergence of Chinese regional clubs was primarily driven by economic initial conditions, preferential development policy, and general government spending, suggesting that policy interventions were effective in overcoming initial disadvantages. In addition, indigenous investments in R&D promoted regional TFP growth (Huang et al., 2019), while knowledge capital and knowledge spillovers promoted productivity in the manufacturing sector at the regional level in China (Scherngell et al., 2014), and competition-friendly industrial policies increased Chinese firms' productivity growth (Aghion et al., 2015). These findings highlight that productivity convergence operated through both capital deepening and technology diffusion channels, with government policy playing a facilitating role.

### 2.2.2 Inland Industrialization and Spatial Rebalancing

The most direct evidence for spatial productivity rebalancing comes from studies examining the inland industrialization process. Deng and Jefferson (2011) shows that Chinese inland regions achieved labour productivity convergence by gaining access to trade, foreign capital, and advanced technologies previously absorbed by coastal regions, while Lemoine et al. (2015) demonstrate that inland industrialization depended more on domestic demand and government support than on international market integration. China's growth model subsequently underwent significant transformation through successive Five-Year Plans. The 12th Plan (2011–2015) marked a shift from investment-heavy export-led growth toward domestic consumption and services, while the 13th Plan (2016–2020) further strengthened social protection and urbanization to foster consumption-driven growth. Complementing these temporal shifts, spatial initiatives like the "Go West" strategy directed substantial investment and infrastructure to western and inland regions, targeting reduction in interregional gaps (Feng et al., 2024). However, the effectiveness of such place-based policies has proven heterogeneous, with growth effects concentrated in localities with stronger initial endowments (Zhu et al., 2025). Our empirical strategy accounts for this heterogeneity through province-level fixed effects within each macro-region and a spatial panel VAR specification that captures cross-province spillovers.

Collectively, these reforms provide the context for evaluating productivity dynamics across Chinese regions. In line with the literature and China's evolving policy framework, three guiding propositions structure our empirical analysis: (1) the persistent force of the Kaldor-Verdoorn effect, ensuring robust increasing returns to scale nationally; (2) a fundamental transition in the drivers of productivity, characterised by the declining role of exports and the ascendant role of government expenditure post-2010; (3) the progressive regionalisation of fiscal effectiveness, whereby government spending has a more potent impact on productivity in less-developed western and central provinces, validating the goals of spatial rebalancing.

# 3 Data and Methods

## 3.1 Data

To analyze how regional demand dynamics and investment intensity influence labour productivity in China, we construct a provincial-level panel dataset spanning 2001–2021, covering the $10^{th}$ through $13^{th}$ five-year plans. The data are drawn from the *China Statistical Yearbook* provided by the Chinese National Bureau of Statistics (see Table 1 for acronyms, description and detailed sources).

| Acronym | Description | Source |
|---|---|---|
| $y$ | GDP at constant prices, in 100 Million Yuan (ref. level 2015) | China Statistical Yearbook |
| $X$ | Exports of goods and services at constant prices, in 100 Million Yuan (ref. level 2015) | China Statistical Yearbook |
| $G$ | Goverment expenditure at constant prices, in 100 Million Yuan (ref. level 2015) | China Statistical Yearbook |
| $Z$ | Autonomous demand at constant prices ($Z = X + G$), in 100 Million Yuan (ref. level 2015) | China Statistical Yearbook |
| $I$ | Gross Fixed Capital Formation at constant prices, in 100 Million Yuan (ref. level 2015) divided by the number of workers (in 10000 persons) | China Statistical Yearbook |
| $p$ | GDP at constant prices, in 100 Million Yuan (ref. level 2015) divided by the number of workers (in 10000 persons) | China Statistical Yearbook |

Table 1: Acronyms, Descriptions and Data Sources

We collected data for exports ($X$), government expenditure ($G$), GDP ($y$), capital deepening (i.e.: Capital Investment per worker-$I$) and labour productivity ($p$).[1] Labour productivity is defined as real Gross Domestic Product (GDP) divided by the number of employed people, while the investment–labour ratio is calculated as real gross fixed capital formation (GFCF) divided by the number of employed people.[2] To proxy regional demand, we use real GDP in the first model, and external demand (exports) and public spending as a proxy for autonomous demand (Barbieri Góes and Deleidi, 2022; Deleidi et al., 2023). All variables are log-transformed prior to estimation. This allows the use of VAR models in levels with variables that may be co-integrated in the long run, in line with Auerbach and Gorodnichenko (2012) and Kilian and Lütkepohl (2017). The regional framework for our analysis is presented in Table (2) which divides China's provinces into Eastern, Central, and Western macroregions that enables examination of spatial heterogeneity in productivity dynamics.

We also divide our sample into two sub-periods (i.e. 2001–2010 and 2011–2020) each spanning two consecutive Five-Year Plans (the 10th and 11th Plans, and the 12th and 13th Plans, respectively). This periodisation is motivated by structural shifts in China's growth model: the first decade is characterised by export-led growth and rapid capital accumulation following WTO accession, while the second reflects a rebalancing towards domestic consumption, the "Go West" strategy's maturation, and decelerating aggregate growth. Splitting the sample at 2010 therefore allows us to examine whether Kaldor–Verdoorn dynamics and the role of autonomous demand shifted across these two distinct policy regimes. The same sub-periods are used to estimate the econometric models in Section (4).

[1]Some variables (government expenditure and employment) have missing values, potentially reducing information content, estimation efficiency (Cameron and Trivedi, 2005), and introducing selection or attrition bias (Baltagi et al., 2013). To address this issue, we first imputed the missing data using the multiple imputation technique developed by Honaker and King (2010), then we performed the Kolmogorov-Smirnoff test (Durbin, 1973) to test the robustness of the imputed dataset. We provide the description of the multiple imputation technique developed by Honaker and King (2010) and the Kolmogorov-Smirnoff Test (Durbin, 1973) in Appendix A.

[2]Since capital stock estimates at the provincial level are not publicly available, we approximate capital accumulation through the growth rate of investment per worker. This proxy is grounded in the convergence properties of investment and capital stock growth, as shown in Freitas and Serrano (2015).

| Eastern regions | Central regions | Western regions |
|---|---|---|
| Beijing | Shanxi | Inner Mongolia |
| Tianjin | Anhui | Guangxi |
| Hebei | Jiangxi | Chongqing |
| Shanghai | Henan | Sichuan |
| Jiangsu | Hubei | Guizhou |
| Zhejiang | Hunan | Yunnan |
| Fujian | Liaoning (NE) | Tibet |
| Shandong | Jilin (NE) | Shaanxi |
| Guangdong | Heilongjiang (NE) | Gansu |
| Hainan | | Qinghai |
| | | Ningxia |
| | | Xinjiang |

Table 2: Chinese provincial-level regions grouped by macro region. Northeast provinces are included under Central, marked with (NE).

Having presented the dataset construction, we now illustrate the spatial and temporal distribution of our outcome variable, labour productivity ($p$), to motivate the regional and temporal dynamics that underpin our empirical analysis. [3]

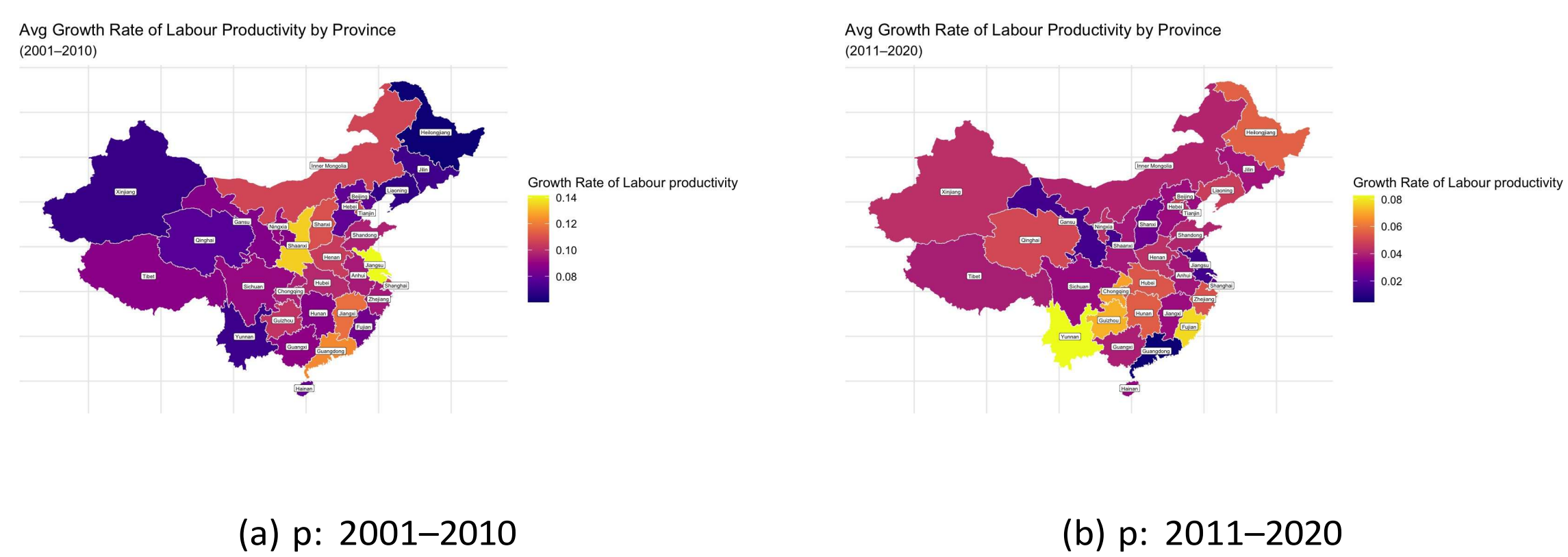

(a) p: 2001–2010

(b) p: 2011–2020

Figure 1: Average Annual Labour Productivity Growth by Province (2001-2010; 2011-2020)

Figure (1) reveals that labour productivity growth exhibits heterogeneous outcomes across China's provinces. Between 2001 and 2010 (see 1a), labour productivity rose fastest in central provinces (Shanxi, Henan, Anhui) and selected eastern regions. Western provinces, though recipients of investment, generally lagged. From 2011–2020 (see Figure 1b), productivity growth concentrated in the southwest. Yunnan and Guizhou led (∼8%), followed by Hainan and Guangxi (∼6–7%). Jiangxi, Fujian, and Guangdong achieved moderate gains (∼4–6%), while the northeast and parts of the northwest (Ningxia, Gansu, Shaanxi) recorded the weakest growth (∼2–3%).

Overall, this evidence confirms substantial regional heterogeneity, supporting the use of fixed-effects models. Temporal division (pre-2010 vs. post-2010) may also reveal shifts in Kaldor-Verdoorn effects as shown in this preliminary analysis of our outcome variable, inspiring non-linear estimations strategy.

[3] Spatial maps for the remaining variables (exports, government expenditure, GDP, and capital per worker) confirm the broad patterns described below and are available upon request.

## 3.2 Methods

To examine the role of demand in determining higher labor productivity at the Chinese regional level, we employ Panel Structural Vector Autoregressive (P-SVAR) modeling. The general P-SVAR framework can be expressed as:

$$B_{0i} z_{i,t} = B_i(L) z_{i,t-n} + w_{i,t} \tag{1}$$

where $B_{0i}$ represents the matrix of contemporaneous coefficients, $z_{i,t-n}$ the vector of endogenous variables, $B_i(L)$ the matrix of lagged coefficients, and $w_{i,t}$ the structural shocks. The structural shocks are identified through theory-based restrictions on $B_{0i}$ (Kilian and Lütkepohl, 2017), drawing on insights from Michl (1985) and Serrano (1995), as well as the identification strategies in Antenucci et al. (2020), Deleidi et al. (2021), Barbieri Góes and Deleidi (2022), Deleidi et al. (2023) and Barbieri Góes et al. (2026).

We estimate two distinct specifications using short-run restrictions, summarized in the systems of equations (2 and 3). In Model 1 (see system of equations 2), GDP ($y_{i,t}$) immediately impacts both the investment-labor ratio ($I_{i,t}$) and labor productivity ($p_{i,t}$); conversely, output does not respond instantaneously to changes in either productivity levels or investment intensity. This recursive structure reflects our theoretical assumption that macroeconomic aggregates drive productivity dynamics in the short run, while productivity feedback effects on output operate with a temporal lag (Barbieri Góes et al., 2026).

$$B_{0i} z_{i,t} = \begin{bmatrix} - & 0 & 0 \\ - & - & 0 \\ - & - & - \end{bmatrix} \begin{bmatrix} y_{i,t} \\ I_{i,t} \\ p_{i,t} \end{bmatrix} \tag{2}$$

In Model 2 (see system of equations 3), we extend the analysis by incorporating exports ($X_{i,t}$) and government spending ($G_{i,t}$). Following Barbieri Góes and Deleidi (2022), we assume that exports are primarily determined by foreign rather than domestic income.[4] This structure is consistent with the view that autonomous demand shapes output first, output then conditions investment decisions, and productivity adjusts last through capital deepening and scale effects. It also aligns with recent P-SVAR applications of the augmented Kaldor–Verdoorn law (Deleidi et al., 2023; Barbieri Góes et al., 2026), where demand shocks are ordered before labor productivity. Second, following established fiscal policy literature (Blanchard and Perotti, 2002; Auerbach and Gorodnichenko, 2012), we assume government spending ($G_{i,t}$) does not respond immediately to output fluctuations due to inherent policy implementation lags and information delays in fiscal decision-making.[5]

[4]While the Chinese economy's size means domestic shocks may eventually affect exports through global spillovers, such transmission likely occurs with significant time lags. This justifies treating exports as contemporaneously unaffected by domestic variables ($G_{i,t}$, $I_{i,t}$, $Y_{i,t}$, $p_{i,t}$).

[5]This identification approach, while originally developed for quarterly data (Blanchard and Perotti, 2002), has gained widespread application in annual panel analyses (Beetsma et al., 2009; Born and Müller, 2012; B´en´etrix and Lane, 2013; Konstantinou and Partheniou, 2021). The assumption of no contemporaneous response of government spending to output is particularly well-grounded in the Chinese institutional context. Under the Budget Law of the People's Republic of China, approved provincial budgets cannot be adjusted without going through formal statutory procedures, and governments are explicitly prohibited from formulating new expenditure policies during budget execution without prior approval (National People's Congress of the People's Republic of China, 2018).

$$B_{0i} z_{i,t} = \begin{bmatrix} - & 0 & 0 & 0 & 0 \\ - & - & 0 & 0 & 0 \\ - & - & - & 0 & 0 \\ - & - & - & - & 0 \\ - & - & - & - & - \end{bmatrix} \begin{bmatrix} X_{i,t} \\ G_{i,t} \\ Y_{i,t} \\ l_{i,t} \\ p_{i,t} \end{bmatrix} \quad (3)$$

After estimating the SVAR models, we calculate impulse response functions (IRFs) to examine how shocks to output ($y_{i,t}$), exports ($X_{i,t}$), government expenditure ($G_{i,t}$) and capital per worker ($l_{i,t}$) influence labor productivity ($p_{i,t}$). The IRFs, presented with 90% confidence intervals, are obtained using a 1000-run moving block bootstrap and cover a 5-year horizon. Following the IRF analysis, we compute the Kaldor-Verdoorn coefficients, as well as the multipliers for exports and government expenditure (in Model 2). Given that the variables are in logs, IRFs can be interpreted as elasticities. To compute multipliers, these elasticities are scaled by the appropriate ex-post conversion factors, translating them into yuan changes in output per yuan change in expenditure.[6] This procedure allows us to recover both the Verdoorn coefficient (the cumulative elasticity of productivity with respect to output) and the demand multipliers associated with exports and government expenditure, thereby quantifying how different autonomous demand components contribute to regional productivity dynamics through their effects on output.

All models are estimated using the full sample period (2001–2021). To assess the robustness of the results, we perform several additional analyses. First, we re-estimate the models excluding the COVID-19 pandemic years (i.e., using data from 2001–2019). Second, we divide the sample into two subperiods 2001–2010, corresponding to the 10th and 11th Five-Year Plans, and 2011–2020, covering the 12th and 13th Five-Year Plans-to examine potential non-linearities over time. Third, we disaggregate the data by geographic regions (East, Central, and West) and re-estimate the models for both the full sample and the two subperiods. All panel estimations incorporate both region and time fixed effects to account for unobserved heterogeneity and common temporal shocks.[7]

# 4 Empirical Findings

This section presents the empirical results by analysing both the dynamic responses captured by the IRFs and the magnitudes of the coefficients derived from cumulative effects. Section 4.1 reports the augmented Kaldor-Verdoorn law from Model 1 by sub-period (Figure 2, Table 3) and macro-region (Figure 3, Table 4). Section 4.2 examines the transition from export-led to domestic-led growth through Model 2 productivity elasticities (Figure 4, Table 13) and output multipliers (Figure 5, Table 14). Section 4.3 analyses spatial rebalancing through regional disaggregation of Model 2 productivity responses (Figure 6, Table 7) and fiscal and export multipliers (Figure 7, Table 8).

## 4.1 Augmented Kaldor-Verdoorn Law

Model 1 (see Figure 2 and Table 3) reveals substantial temporal heterogeneity in Kaldor-Verdoorn dynamics. Output shocks ($y \rightarrow p$) exhibit stable positive effects across all periods, with cumulative coefficients ranging from 0.65 to 0.76, while capital per worker shocks ($I \rightarrow p$) weaken considerably in 2011-2020, averaging just 0.15 compared to 0.38 in the full sample.

[6]We present cumulative multipliers, which measure the total output response relative to the total change in export ($X$) and government expenditure ($G$) over time. These multipliers, considered the most reliable metric for assessing the macroeconomic impact of fiscal shocks (Spilimbergo et al., 2009; Ramey and Zubairy, 2018), are obtained by dividing the cumulative output variation by the cumulative expenditure change.

[7]As a further robustness test, to account for possible spillover effects, we estimate a spatial panel VAR (SPVAR) model, the results of which are reported in Appendix B.

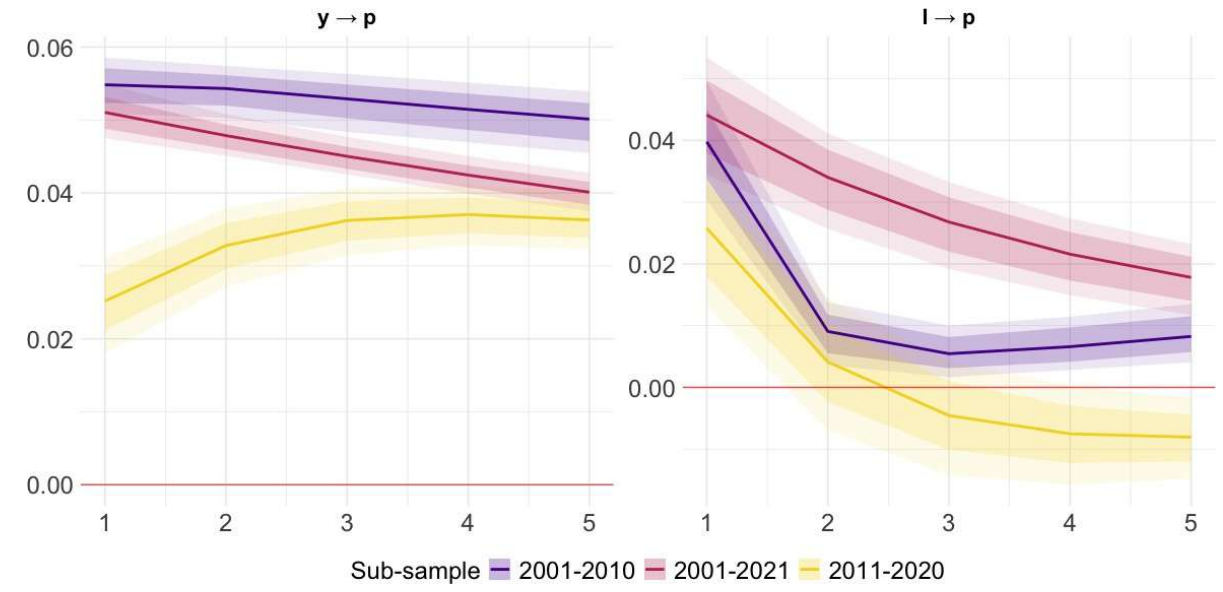

Figure 2: **IRFs, Model 1 by sub-period (2001-2021; 2001-2010; 2011-2020):** Figures display IRFs of labour productivity to output ($y$) and investment per worker ($I$) shocks. Years on x-axis. Shaded areas denote 90% and 68% confidence bands calculated through m.b. bootstrapping (1000 runs).

| | 1 | 2 | 3 | 4 | 5 | Avg |
|---|---|---|---|---|---|---|
| | | | 2001-2021 | | | |
| $y$ | **0.69** | **0.68** | **0.68** | **0.68** | **0.67** | 0.68 |
| $I$ | **0.38** | **0.38** | **0.37** | **0.37** | **0.37** | 0.38 |
| | | | 2001-2010 | | | |
| $y$ | **0.66** | **0.67** | **0.68** | **0.68** | **0.68** | 0.67 |
| $I$ | **0.41** | **0.33** | **0.28** | **0.26** | **0.26** | 0.31 |
| | | | 2011-2020 | | | |
| $y$ | **0.59** | **0.70** | **0.78** | **0.84** | **0.89** | 0.76 |
| $I$ | **0.24** | 0.20 | 0.16 | 0.11 | 0.06 | 0.15 |

Table 3: **Cumulative effects estimated for Models 1 by sub-period (2001–2021; 2001–2010; 2011–2020).** Statistically significant values in bold (90% c.i.). The average effect is estimated across 5 years.

Figure (3) and Table (4) show results by macro-region. Over the full sample, output shocks raise productivity similarly across all regions (0.67-0.68), yet the East exhibits a far stronger capital per worker coefficient (0.53 versus 0.35 and 0.26). In 2001-2010, the East maintains the highest capital-productivity link (0.41) while all three macroregions display similarly high Verdoorn coefficients (0.66-0.68), signalling widespread increasing returns. The pattern reverses after 2011. Verdoorn coefficients surge inland (0.82 in Central and 0.78 in the West versus 0.63 in the East), while the East's capital per worker effect nearly halves.[8]

[8]The negative capital per worker coefficient in Central China during 2011-2020 (averaging -0.23) reflects a specific historical conjuncture in which the composition of investment, rather than its volume, determined productivity outcomes. When the northeastern provinces (Liaoning, Jilin, Heilongjiang) are excluded from the Central region, the coefficient loses statistical significance, suggesting that severe industrial restructuring and demographic decline in these provinces compressed the productivity impact of new capital during this period. Under these conditions, measured investment per worker proxies for capital deepening in low-productivity sectors rather than productive capacity expansion, generating a negative relationship with labour productivity that is regime-specific rather than a general feature.

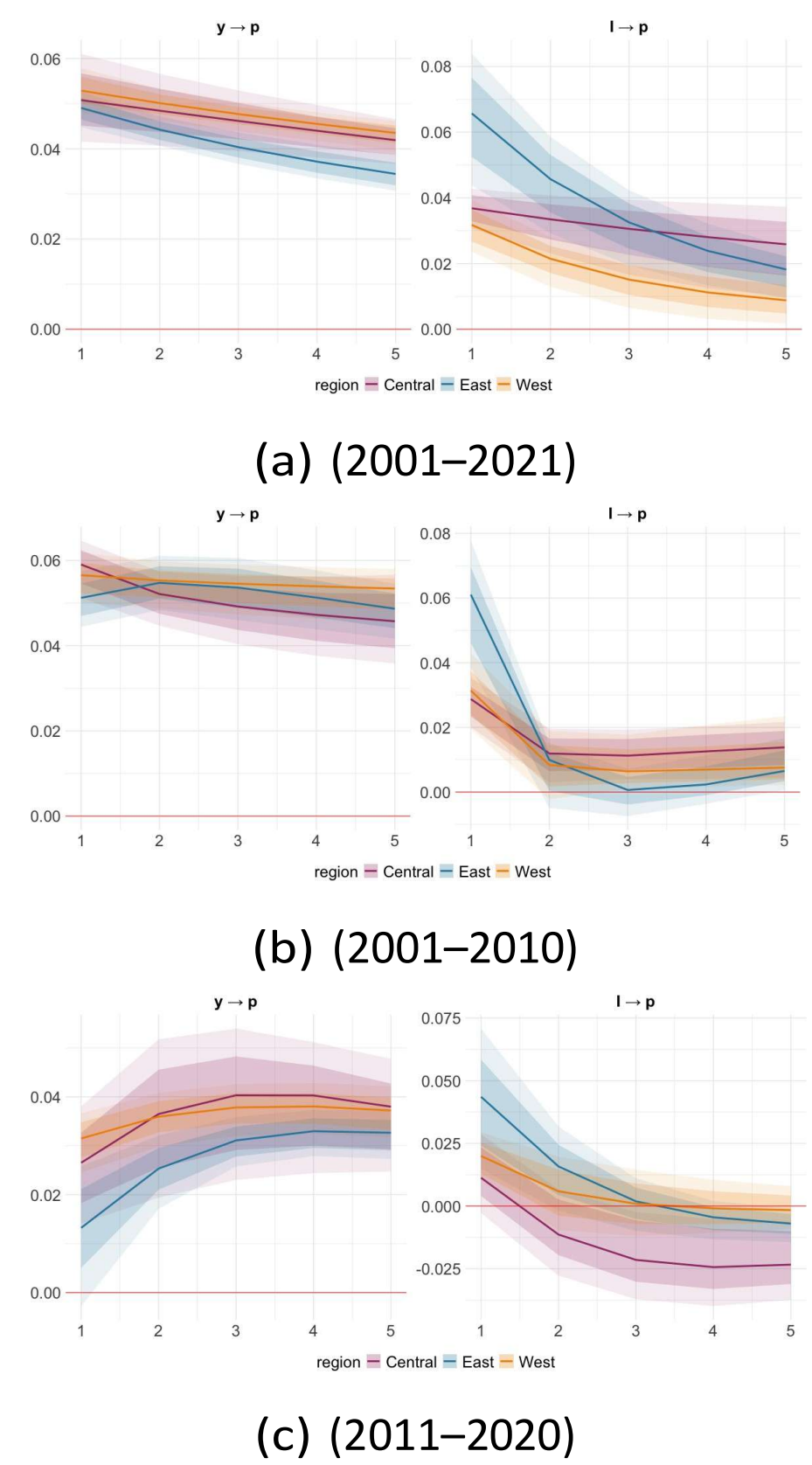

(a) (2001–2021)

(b) (2001–2010)

(c) (2011–2020)

Figure 3: **IRFs, Model 1 by macro-region (East; Central; West) and sub-period (2001-2021; 2001-2010; 2011-2020):** Figures display IRFs of labour productivity to output ($y$) and investment per worker ($I$) shocks. Years on x-axis. Shaded areas denote 90% and 68% confidence bands calculated through m.b. bootstrapping (1000 runs).

| | | 1 | 2 | 3 | 4 | 5 | Avg |
|---|---|---|---|---|---|---|---|
| | | | | **2001-2021** | | | |
| **East** | $y$ | **0.70** | **0.68** | **0.67** | **0.66** | **0.65** | 0.67 |
| | $I$ | **0.58** | **0.55** | **0.52** | **0.50** | **0.48** | 0.53 |
| **Central** | $y$ | **0.67** | **0.67** | **0.68** | **0.68** | **0.68** | 0.68 |
| | $I$ | **0.34** | **0.34** | **0.35** | **0.36** | **0.37** | 0.35 |
| **West** | $y$ | **0.69** | **0.69** | **0.68** | **0.68** | **0.68** | 0.68 |
| | $I$ | **0.26** | **0.26** | **0.26** | **0.26** | **0.27** | 0.26 |
| | | | | **2001-2010** | | | |
| **East** | $y$ | **0.63** | **0.67** | **0.69** | **0.70** | **0.70** | 0.68 |
| | $I$ | **0.61** | **0.46** | **0.37** | **0.31** | **0.29** | 0.41 |
| **Central** | $y$ | **0.72** | **0.69** | **0.68** | **0.67** | **0.66** | 0.68 |
| | $I$ | **0.34** | **0.30** | **0.29** | **0.30** | **0.31** | 0.31 |
| **West** | $y$ | **0.66** | **0.66** | **0.66** | **0.66** | **0.66** | 0.66 |
| | I | **0.31** | **0.26** | 0.24 | **0.23** | **0.24** | 0.26 |
| | | | | **2011-2020** | | | |
| **East** | $y$ | 0.36 | **0.54** | **0.67** | **0.76** | **0.82** | 0.63 |
| | $I$ | **0.43** | 0.39 | 0.35 | 0.31 | 0.27 | 0.35 |
| **Central** | $y$ | **0.62** | **0.75** | **0.85** | **0.92** | **0.98** | 0.82 |
| | $I$ | 0.12 | 0.00 | -0.16 | **-0.39** | **-0.72** | -0.23 |
| **West** | $y$ | **0.67** | **0.74** | **0.79** | **0.83** | **0.86** | 0.78 |
| | $I$ | **0.17** | 0.17 | 0.16 | 0.16 | 0.15 | 0.16 |

Table 4: **Cumulative effects estimated for Models 1 by macro-region (East; Central; West) and sub-period (2001-2021; 2001-2010; 2011-2020).** Statistically significant values in bold (90% c.i.). The average effect is estimated across 5 years.

## 4.2 From Export-led growth to domestic-led growth

Model 2 (Tables 13 and 14) provides the most comprehensive view by incorporating autonomous demand components ($X$ and $G$) alongside output ($Y$) and capital per worker ($I$). Output coefficients remain robust across all specifications, averaging 0.80 in the full sample (0.77 in 2001-2010, 0.66 in 2011-2020). Export effects on productivity decline dramatically from 0.20 (2001-2010) to 0.01 (2011-2020), while government expenditure effects strengthen from 0.31 to 0.40. This aligns with broader structural changes documented in the literature. The export slowdown led local party secretaries to increase social spending (Tombe and Zhu, 2019), while Chinese employment growth was driven by domestic rather than export demand (Los et al., 2015). Capital per worker coefficients decline from 0.34 to 0.17, mirroring Model 1 results.

Transforming elasticities into multipliers reveals a striking temporal pattern. Export multipliers averaged 1.11 in 2001-2010, indicating that each yuan of exports generated approximately 1.11 yuan of output, collapsing to just 0.10 in 2011-2020. Government expenditure multipliers demonstrate remarkable consistency throughout, with an average fiscal multiplier of 2.51 over the full sample (each yuan of government spending generating approximately 2.5 yuan of output) declining only modestly from 3.34 in 2001-2010 to 2.47 in 2011-2020. These findings complement recent evidence on fiscal multipliers in China (He and Yu, 2025) by identifying the Kaldor-Verdoorn productivity channel as the key transmission mechanism linking fiscal shocks to long-run labour productivity outcomes.

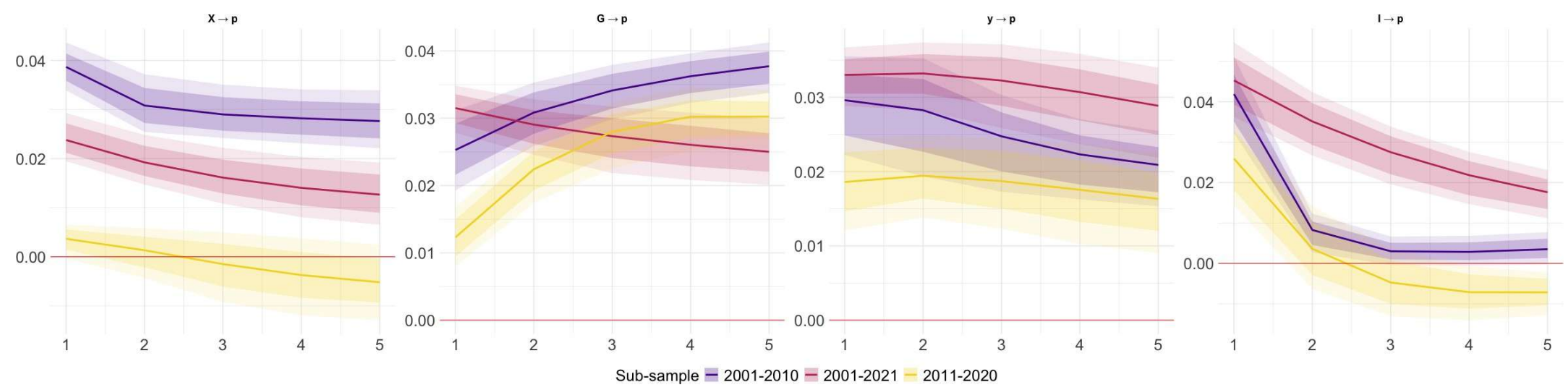

Figure 4: **IRFs, Model 2 by sub-period (2001-2021; 2001-2010; 2011-2020):** Figures display IRFs of labour productivity (*p*) to export (*X*), government expenditure (*G*), output (*y*), and investment per worker (*I*) shocks. Years on x-axis. Shaded areas denote 90% and 68% confidence bands calculated through m.b. bootstrapping (1000 runs).

| | 1 | 2 | 3 | 4 | 5 | Avg |
|---|---|---|---|---|---|---|
| | | | **2001-2021** | | | |
| *X* | **0.10** | **0.10** | **0.10** | **0.10** | **0.10** | 0.10 |
| *G* | **0.30** | **0.31** | **0.32** | **0.32** | **0.33** | 0.32 |
| *y* | **0.74** | **0.78** | **0.81** | **0.82** | **0.83** | 0.80 |
| *I* | **0.40** | **0.40** | **0.40** | **0.39** | **0.39** | 0.40 |
| | | | **2001-2010** | | | |
| *X* | **0.16** | **0.18** | **0.20** | **0.23** | **0.26** | 0.20 |
| *G* | **0.24** | **0.29** | **0.32** | **0.35** | **0.36** | 0.31 |
| *y* | **0.72** | **0.78** | **0.79** | **0.79** | **0.79** | 0.77 |
| *I* | **0.45** | **0.36** | 0.31 | 0.28 | 0.27 | 0.34 |
| | | | **2011-2020** | | | |
| *X* | 0.02 | 0.01 | 0.01 | 0.00 | -0.01 | 0.01 |
| *G* | **0.21** | **0.32** | **0.41** | **0.50** | **0.57** | 0.40 |
| *y* | **0.56** | **0.62** | **0.66** | **0.70** | **0.73** | 0.66 |
| *I* | **0.25** | 0.21 | 0.17 | 0.12 | 0.08 | 0.17 |

Table 5: **Cumulative effects estimated for Models 2 by sub-period (2001-2021; 2001-2010; 2011-2020).** Statistically significant values in bold (90% c.i.). The average effect is estimated across 5 years.

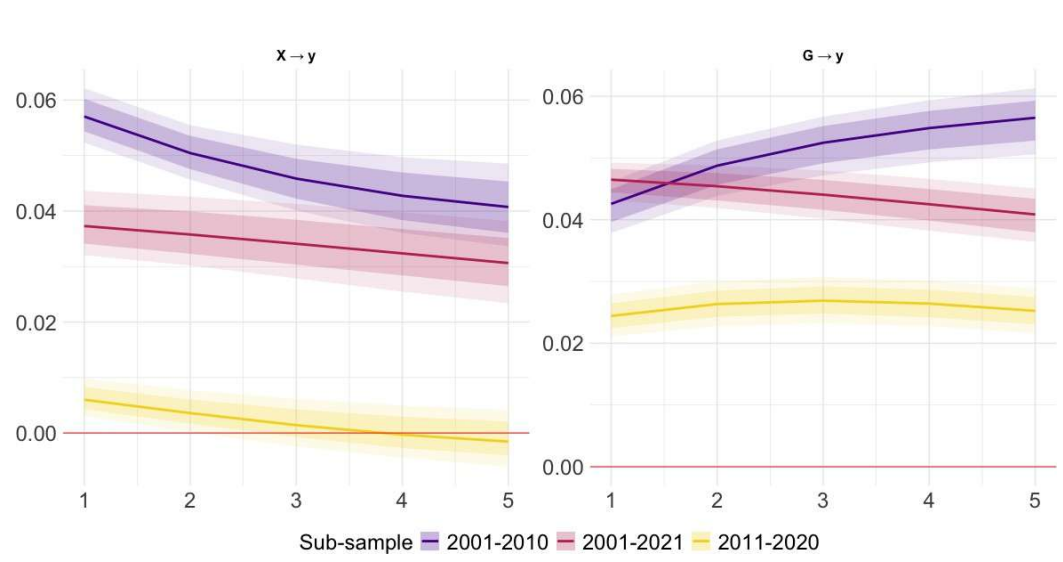

Figure 5: **IRFs, Model 2 by sub-period (2001-2021; 2001-2010; 2011-2020):** Figures display IRFs of output to export (*X*) and government expenditure (*G*) shocks. Years on x-axis. Shaded areas denote 90% and 68% confidence bands calculated through m.b. bootstrapping (1000 runs).

| | 1 | 2 | 3 | 4 | 5 | Avg |
|---|---|---|---|---|---|---|
| | | | **2001-2021** | | | |
| *X* | **0.71** | **0.75** | **0.79** | **0.83** | **0.87** | 0.79 |
| *G* | **2.29** | **2.42** | **2.53** | **2.62** | **2.71** | 2.51 |
| | | | **2001-2010** | | | |
| *X* | **0.83** | **0.96** | **1.10** | **1.25** | **1.40** | 1.11 |
| *G* | **2.75** | **3.13** | **3.42** | **3.63** | **3.79** | 3.34 |
| | | | **2011-2020** | | | |
| *X* | **0.14** | **0.13** | **0.11** | **0.09** | **0.06** | 0.10 |
| *G* | **2.00** | **2.26** | **2.50** | **2.70** | **2.89** | 2.47 |

Table 6: **Cumulative multipliers of output to export (*X*) and government expenditure (*G*) shocks, Model 2 by sub-period (2001–2021; 2001–2010; 2011–2020):** Statistically significant values (90% confidence bands) are in bold. Cumulative multipliers are computed as the cumulative change in output over the cumulative change in the respective variable across horizons up to 5 years. The average multiplier represents the mean cumulative effect over the 5-year horizon.

### 4.3 Spatial Rebalancing

Analyzing Model 2 by macro-region and sub-periods (see Figures 6 and 7 and Tables 7 and 8), several interesting patterns emerge. Eastern regions maintain their position as China's export powerhouse throughout all time periods, with export productivity coefficients averaging 0.29 over the full sample, nearly triple that of Central China (0.09) and more than four times that of the West (0.07). Eastern export coefficients decline notably from 0.29 in 2001-2010 to 0.11 in 2011-2020, a 62% reduction paralleling the broader collapse in export multipliers observed nationally. The paradox whereby the East shows the highest export-productivity coefficients but the lowest export multipliers (1.15 versus 3.01 in Central China and 1.46 in the West) reflects the region's already high export intensity and mature industrial structure, consistent with Blecker's (2000) findings on diminishing returns in export-intensive economies. Across all regions, export multipliers collapse dramatically in 2011-2020 (0.17 in the East, 1.57 in Central China, and 0.15 in the West), reflecting the structural transformation away from export-led growth.

With respect to government expenditure, redistributive policies in favour of West and Central China successfully closed the gap in fiscal multipliers while strengthening the Verdoorn coefficients. Central and Western China demonstrate stronger $G \rightarrow p$ effects than Eastern regions (0.42 and 0.30 versus 0.22 over the full sample), while fiscal multipliers follow the opposite pattern (2.90 in the East versus 2.44 and 1.71 in Central and Western regions respectively). The $G \rightarrow p$ relationship strengthens across all regions in 2011-2020, with Western provinces showing the most dramatic improvement (57.6% growth from 0.33 to 0.52). Conversely, fiscal multipliers decline across all macro-regions but converge substantially by 2011-2020 (2.49 in the East versus 2.48 in the West), validating the effectiveness of China's redistributive fiscal policies and the "Go West" strategy in promoting productivity growth in historically disadvantaged regions.

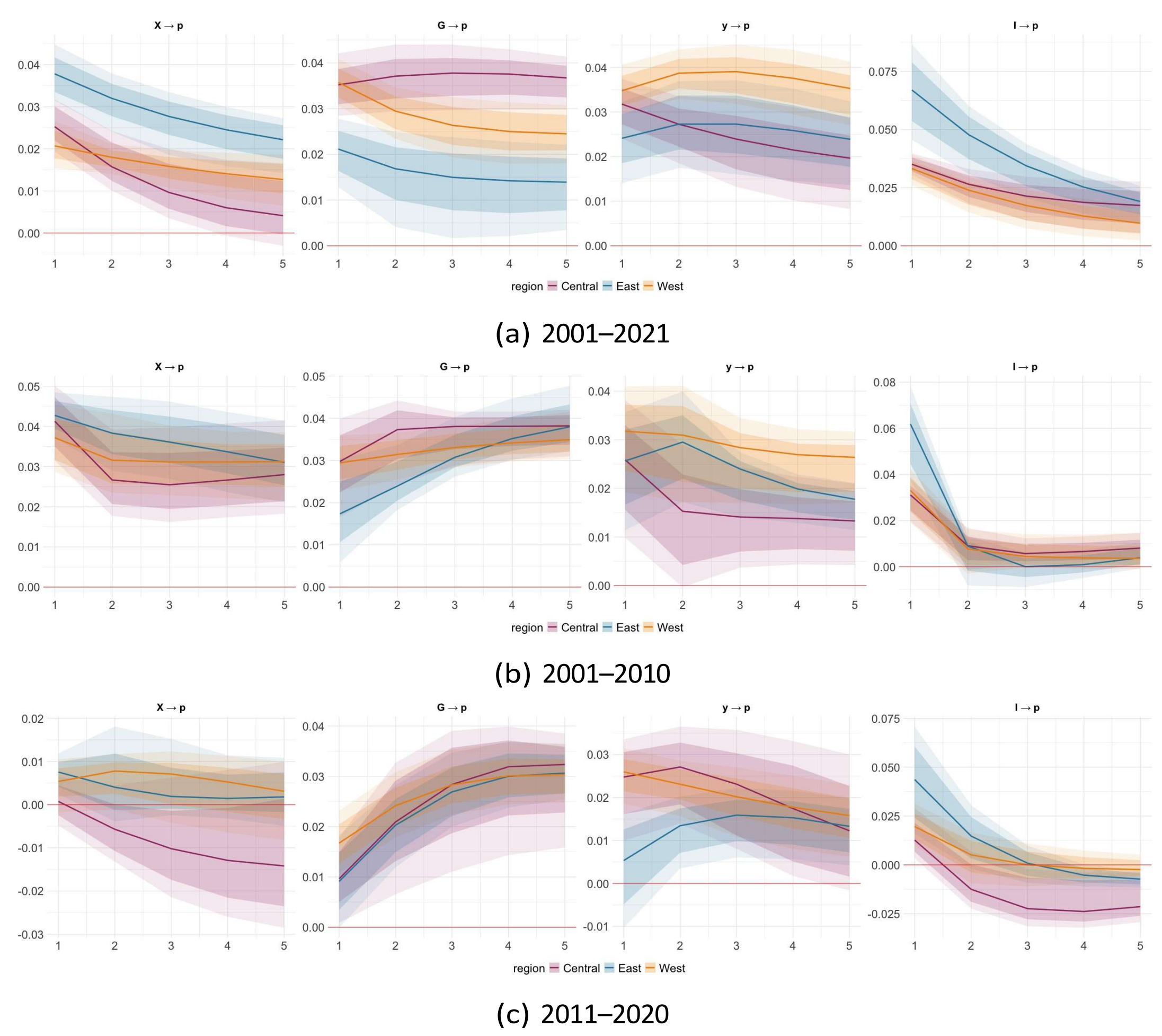

Figure 6: **IRFs, Model 2 by macro-region (East; Central; West) and sub-period (2001-2021; 2001-2010; 2011-2020):** Figures display IRFs of labour productivity to export ($X$), government expenditure ($G$), output ($Y$), and investment per worker ($I$) shocks. Years on x-axis. Shaded areas denote 90% and 68% confidence bands calculated through m.b. bootstrapping (1000 runs).

| | | 1 | 2 | 3 | 4 | 5 | Avg |
|---|---|---|---|---|---|---|---|
| | | | | 2001-2021 | | | |
| East | *X* | **0.26** | **0.27** | **0.29** | **0.31** | **0.33** | 0.29 |
| | *G* | **0.22** | **0.21** | 0.22 | 0.22 | **0.23** | 0.22 |
| | *y* | **0.69** | **0.75** | **0.78** | **0.79** | **0.79** | 0.76 |
| | *I* | **0.60** | **0.57** | **0.54** | **0.52** | **0.50** | 0.54 |
| Central | *X* | **0.11** | **0.10** | **0.08** | 0.07 | 0.07 | 0.09 |
| | *G* | **0.37** | **0.40** | **0.43** | **0.45** | **0.47** | 0.42 |
| | *y* | **0.79** | **0.82** | **0.84** | **0.85** | **0.85** | 0.83 |
| | *I* | **0.33** | **0.32** | **0.31** | **0.30** | **0.31** | 0.31 |
| West | *X* | **0.07** | **0.07** | **0.07** | **0.07** | **0.07** | 0.07 |
| | *G* | **0.32** | **0.30** | **0.29** | **0.29** | **0.30** | 0.30 |
| | *y* | **0.72** | **0.80** | **0.85** | **0.88** | **0.89** | 0.83 |
| | *I* | **0.28** | **0.29** | **0.30** | **0.30** | **0.31** | 0.30 |
| | | | | 2001-2010 | | | |
| East | *X* | **0.23** | **0.26** | **0.29** | **0.32** | **0.36** | 0.29 |
| | *G* | **0.18** | **0.23** | **0.28** | **0.32** | **0.35** | 0.27 |
| | *y* | **0.84** | **1.00** | **1.03** | **1.01** | **0.99** | 0.97 |
| | *I* | **0.64** | 0.49 | 0.39 | 0.33 | 0.29 | 0.43 |
| Central | *X* | **0.16** | **0.16** | **0.17** | **0.19** | **0.21** | 0.18 |
| | *G* | **0.30** | **0.36** | **0.39** | **0.41** | **0.42** | 0.38 |
| | *y* | **0.69** | 0.66 | **0.66** | **0.67** | **0.68** | 0.67 |
| | *I* | **0.39** | 0.33 | 0.29 | 0.27 | 0.27 | 0.31 |
| West | *X* | **0.14** | **0.16** | **0.20** | **0.24** | **0.28** | 0.21 |
| | *G* | **0.28** | **0.31** | **0.33** | **0.35** | **0.37** | 0.33 |
| | *y* | **0.68** | **0.71** | **0.72** | **0.72** | **0.71** | 0.71 |
| | *I* | **0.35** | 0.29 | 0.27 | 0.27 | 0.27 | 0.29 |
| | | | | 2011-2020 | | | |
| East | *X* | 0.08 | 0.11 | 0.12 | 0.12 | 0.14 | 0.11 |
| | *G* | 0.13 | **0.22** | **0.30** | **0.37** | **0.43** | 0.29 |
| | *y* | 0.20 | 0.38 | **0.52** | **0.61** | **0.68** | 0.48 |
| | *I* | **0.45** | 0.41 | 0.37 | 0.33 | 0.29 | 0.37 |
| Central | *X* | 0.01 | -0.02 | -0.04 | -0.07 | -0.09 | -0.04 |
| | *G* | 0.18 | **0.31** | **0.43** | **0.53** | **0.61** | 0.41 |
| | *y* | **0.79** | **0.92** | **1.00** | **1.05** | 1.06 | 0.97 |
| | *I* | 0.13 | 0.00 | -0.17 | -0.40 | -0.73 | -0.23 |
| West | *X* | 0.02 | 0.03 | 0.03 | 0.03 | 0.03 | 0.03 |
| | *G* | **0.33** | **0.44** | **0.54** | **0.62** | **0.69** | 0.52 |
| | *y* | **0.75** | **0.80** | **0.82** | **0.84** | **0.86** | 0.81 |
| | *I* | **0.18** | 0.17 | 0.17 | 0.15 | 0.14 | 0.16 |

Table 7: Cumulative effects estimated for Models 2 by macro-region (East; Central; West) and sub-period (2001-2021; 2001-2010; 2011-2020). Statistically significant values in bold (90% c.i.). The average effect is estimated across 5 years.

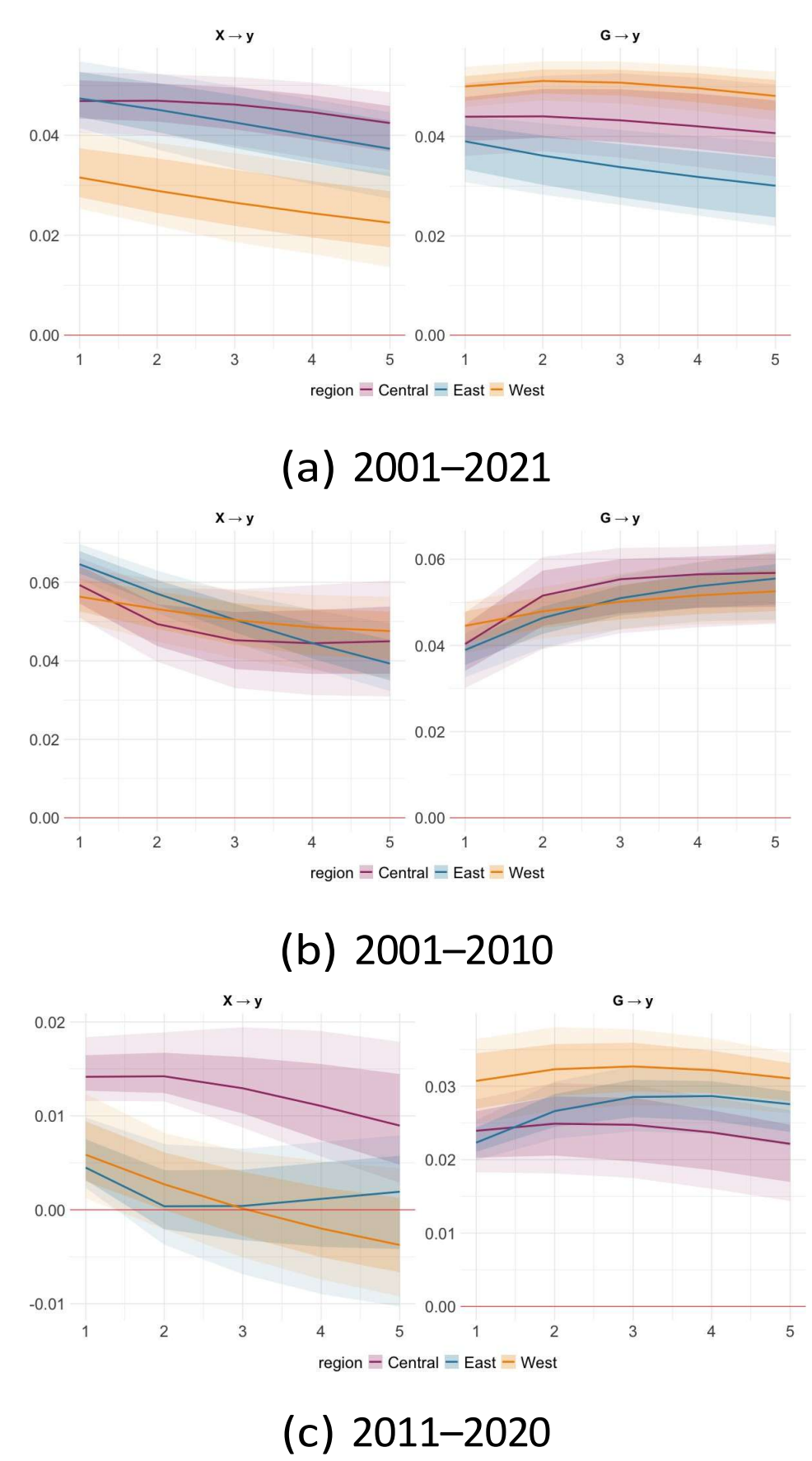

Figure 7: **IRFs, Model 2 by macro-region (East; Central; West) and sub-period (2001-2021; 2001-2010; 2011-2020):** Figures display IRFs of output to export ($X$) and government expenditure ($G$) shocks. Years on x-axis. Shaded areas denote 90% and 68% confidence bands calculated through m.b. bootstrapping (1000 runs).

| | | 1 | 2 | 3 | 4 | 5 | Avg |
|---|---|---|---|---|---|---|---|
| | | | | **2001-2021** | | | |
| **East** | $X$ | **0.92** | **1.03** | **1.15** | **1.26** | **1.37** | 1.15 |
| | $G$ | **2.61** | **2.77** | **2.92** | **3.05** | **3.16** | 2.90 |
| **Central** | $X$ | **2.73** | **2.87** | **3.01** | **3.16** | **3.30** | 3.01 |
| | $G$ | **2.20** | **2.34** | **2.46** | **2.55** | **2.63** | 2.44 |
| **West** | $X$ | **1.36** | **1.41** | **1.46** | **1.51** | **1.56** | 1.46 |
| | $G$ | **1.56** | **1.65** | **1.72** | **1.79** | **1.85** | 1.71 |
| | | | | **2001-2010** | | | |
| **East** | $X$ | **0.76** | **0.84** | **0.93** | **1.01** | **1.10** | 0.93 |
| | $G$ | **3.41** | **4.01** | **4.45** | **4.76** | **4.99** | 4.32 |
| **Central** | $X$ | **2.78** | **3.07** | **3.41** | **3.80** | **4.22** | 3.46 |
| | $G$ | **2.52** | **3.09** | **3.43** | **3.63** | **3.77** | 3.29 |
| **West** | $X$ | **3.31** | **4.10** | **5.01** | **5.99** | **6.99** | 5.08 |
| | $G$ | **1.87** | **2.08** | **2.25** | **2.39** | **2.49** | 2.22 |
| | | | | **2011-2020** | | | |
| **East** | $X$ | **0.16** | 0.14 | 0.14 | 0.17 | 0.22 | 0.17 |
| | $G$ | **1.90** | **2.23** | **2.52** | **2.78** | **3.00** | 2.49 |
| **Central** | $X$ | **1.38** | **1.50** | **1.60** | **1.67** | **1.71** | 1.57 |
| | $G$ | **2.03** | **2.26** | **2.45** | **2.59** | **2.69** | 2.40 |
| **West** | $X$ | 0.25 | 0.21 | 0.15 | 0.10 | 0.04 | 0.15 |
| | $G$ | **2.04** | **2.28** | **2.50** | **2.70** | **2.87** | 2.48 |

Table 8: Cumulative multipliers of output to export ($X$) and government expenditure ($G$) shocks, Model 2 by macro-region (East; Central; West) sub-period (2001–2021; 2001–2010; 2011–2020): Statistically significant values (90% confidence bands) are in bold. Cumulative multipliers are computed as the cumulative change in output over the cumulative change in the respective variable across horizons up to 5 years. The average multiplier represents the mean cumulative effect over the 5-year horizon.

# 5 Conclusion

This study complements the growth regime framework (Baccaro and Pontusson, 2016; Hassel and Palier, 2021) with a demand-led perspective linking regime composition to productivity dynamics through the Kaldor-Verdoorn law, contributing to the emerging literature on subnational growth regimes (Di Carlo et al., 2024; Regan and Blyth, 2025) while extending it to a large transitional economy. Using P-SVAR modelling on Chinese provincial data from 2001 to 2021, we document three core findings.

First, the Kaldor-Verdoorn coefficient remains robust across all regions and time periods (0.65 to 0.76), confirming persistent increasing returns to scale throughout China's structural transformation. Second, China's transition from export-led to domestically-oriented growth is empirically detectable through the Kaldor-Verdoorn mechanism: export effects on productivity collapse from 0.20 in 2001 to 2010 to 0.01 in 2011 to 2020, while government expenditure effects strengthen from 0.31 to 0.40, with fiscal multipliers remaining high throughout (averaging 2.51 yuan per yuan of spending). Third, the post-2011 spatial reversal, wherein central and western regions achieve stronger Verdoorn coefficients (0.82 and 0.78 versus 0.63 in the East) while fiscal multipliers converge between coastal and inland areas, validates the effectiveness of China's redistributive fiscal policies in narrowing regional productivity gaps, extending Lemoine et al. (2015) and Deng and Jefferson (2011) by demonstrating that convergence operates through demand-driven rather than purely technological channels, and contradicting Liu et al. (2022) by showing that redistributive policies enhanced productivity in disadvantaged regions without undermining national performance.

These findings offer important policy guidance for emerging economies pursuing growth rebalancing. Export-led strategies face diminishing returns in mature phases of development, while domestically-oriented fiscal policies can remain robust engines of productivity growth. China's experience demonstrates that strategic transition toward fiscal-anchored domestic demand, combined with spatial targeting toward lagging regions, offers a viable pathway for reducing regional productivity gaps in large heterogeneous economies. Future research should examine whether these demand-driven productivity mechanisms extend to other large emerging economies undergoing similar growth regime transitions, and whether subnational fiscal targeting can be further refined to address within-region heterogeneity (Zhu et al., 2025).

# Appendix A Multiple imputation, diagnostics for imputed dataset and imputation results

## Appendix A.1 Multiple imputation

In multiple imputation, $m$ values are imputed for each missing cell in the data matrix and then $m$ "completed" datasets are created where the observed values are the same, but the missing values are filled in with different imputations (King et al., 2001). The multiple imputation technique developed by Honaker and King (2010) uses a predictive model that incorporates all available information in the observed data together along with any prior knowledge. Honaker and King (2010) describe the predictive model by denoting $D$ as a vector of $p$ variables, containing both explanatory and prediction variables. Vector $D$ is partitioned into observed and missing elements:

$D = (D^{observed}, D^{missing})$

then, $M$ is defined as the missingness indicator matrix in which each element $x$ is

$$x = \begin{cases} 0, & \text{for } D^{\text{observed}}, \\ 1, & \text{for } D^{\text{missing}}. \end{cases}$$

Moreover, multiple imputation models are based on the missing at random (MAR) assumption in which the missingness indicator matrix $M$ can only be predicted by

$D^{observed}$: $p(M \mid D) = p(M \mid D^{\text{observed}})$

Multiple imputation provides efficient and unbiased inference under reasonable distributions of missing data, while taking into account its intrinsic uncertainty (Van Buuren, 2018). Consequently, the (Honaker and King, 2010) algorithm takes m bootstrap samples and applies an expectation-maximization with bootstrap algorithm to each sample. The expectation-maximization produces the point estimates of the mean ($\mu$) and variance ($\Sigma$) of the ($p$) variables (Dempster et al., 1977); then, the process of drawing ($\mu$) and ($\Sigma$) is performed by a bootstrapping algorithm to simulate the estimation uncertainty, ensuring that the m estimates of mean and variance are different in each draw. The Honaker and King (2010) algorithm draws m samples of size n with replacement from data ($D$), which is assumed to be distributed as a multivariate normal

$D \sim \mathrm{N}(\mu, \Sigma)$

Then, for each set of estimates, the original sample units are used to impute the missing observations in their original position, resulting in m imputed data. We set the algorithm to perform a non-parametric bootstrap by setting the maximum resample equal to 1000. We then add a ridge prior equal to 5% of the number of observations to increase the numerical stability of the EM algorithm, and thus avoiding different EM chain lengths (i.e.: the number of iterations required to reach convergence) for each imputation. To clarify, the ridge prior improves the convergence of each multiple imputation, allowing the EM algorithm to easily find the maximum by shrinking the covariances towards zero without changing the means and variances for datasets with high missingness, small observations or large correlations between variables. In addition, both past and future values of a variable may be correlated with the present value of the variables of the imputation model. Therefore, in order to improve the predictive power of the imputation models and thus the accuracy of the imputation results, we introduce lags and leads of intangible assets. Lastly, by computing multiple imputation, we have been able to produce a dataset without missing data.

## Appendix A.2 Diagnostics for imputed dataset

The step following the implementation of the multiple imputation algorithm is to assess the accuracy of the data of the imputation process, therefore we need a rule to test the quality, plausibility and reliability of the imputed dataset. We therefore hypothesize that the observed distribution of a variable should not differ from the imputed distribution of the same variable. On the one hand, we need to rule out the possibility that we would observe a completely different distribution of the variable if data were not missing, and, on the other hand, we need to ensure that the information contained in the missing data is consistent with that provided by the observed data. Following Abayomi et al. (2008), this diagnostic exercise is carried out using the two-samples Kolmogorov–Smirnoff (KS) test. This is a non-parametric test of equality that allows us to compare two samples (i.e.: the observed and the imputed data related to a single variable) by quantifying the distance between the empirical distribution of these two given samples (Abayomi et al., 2008). It follows that the KS test compares the empirical distributions of the observed and the imputed datasets, allowing us to validate or reject the imputed data. The two samples KS test works as follows. We first compute the distance among the observed and imputed data distribution, hence it can be written as:

$D_{(n,m)} = \max \; F_n^{(obs)} - F_m^{(imp)}$

where $D_{(n,m)}$ represents the distance among the cumulative distribution functions $F$, where $F_n^{(obs)}$ and $F_m^{(imp)}$ are the empirical distribution functions of the observed and the imputed variable, respectively. The null hypothesis of the two-samples Kolmogorov-Smirnoff Test states that the true distribution of the observed variable is equal to the distribution of the imputed one, and it can be written as:

$H_0 : F_n^{(obs)} = F_m^{(imp)}$

To rely on our imputed dataset, the KS test should not indicate a statistically significant difference between observed and imputed values.

## Appendix A.3 Imputation results

Table 9 reports the imputation results for the variables Government expenditure, and Number of workers.

| | Government expenditure | | Number of workers | |
|---|---|---|---|---|
| | Statistic | p-value | Statistic | p-value |
| Anhui | 0.1681 | 0.8967 | 0.3809 | 0.1384 |
| Beijing | 0.1317 | 0.9669 | 0.2476 | 0.5431 |
| Chongqing | 0.1793 | 0.8374 | 0.3333 | 0.2542 |
| Fujian | 0.1793 | 0.8314 | 0.2952 | 0.3273 |
| Gansu | 0.1905 | 0.7701 | 0.3333 | 0.2188 |
| Guangdong | 0.1905 | 0.7845 | 0.2381 | 0.6355 |
| Guangxi | 0.1681 | 0.8910 | 0.2857 | 0.3992 |
| Guizhou | 0.1457 | 0.9112 | 0.3333 | 0.2188 |
| Hainan | 0.1793 | 0.8375 | 0.3571 | 0.1927 |
| Hebei | 0.1905 | 0.7845 | 0.3333 | 0.2554 |
| Heilongjiang | 0.1793 | 0.8422 | 0.2619 | 0.5420 |
| Henan | 0.1569 | 0.9120 | 0.3333 | 0.2469 |
| Hubei | 0.1681 | 0.8863 | 0.3333 | 0.2469 |
| Hunan | 0.1681 | 0.8876 | 0.3809 | 0.1450 |
| Inner Mongolia | 0.1457 | 0.9515 | 0.3333 | 0.2576 |
| Jiangsu | 0.1905 | 0.7825 | 0.1714 | 0.9021 |
| Jiangxi | 0.1905 | 0.7937 | 0.2857 | 0.3815 |
| Jilin | 0.1905 | 0.7949 | 0.3333 | 0.2405 |
| Liaoning | 0.1905 | 0.7748 | 0.2381 | 0.5854 |
| Ningxia | 0.1569 | 0.9191 | 0.3333 | 0.2470 |
| Qinghai | 0.1568 | 0.8939 | 0.3333 | 0.2347 |
| Shaanxi | 0.1681 | 0.8804 | 0.3333 | 0.2147 |
| Shandong | 0.1681 | 0.8941 | 0.3333 | 0.2412 |
| Shanghai | 0.1905 | 0.7949 | 0.2381 | 0.6590 |
| Shanxi | 0.1681 | 0.8917 | 0.3142 | 0.2614 |
| Sichuan | 0.1905 | 0.8038 | 0.3333 | 0.2549 |
| Tianjin | 0.1681 | 0.8637 | 0.2476 | 0.5479 |
| Tibet | 0.1905 | 0.7846 | 0.3095 | 0.3150 |
| Xinjiang | 0.1429 | 0.9451 | 0.3571 | 0.1866 |
| Yunnan | 0.1681 | 0.8794 | 0.3333 | 0.2528 |
| Zhejiang | 0.1429 | 0.9387 | 0.2952 | 0.3468 |

Table 9: Kolmogorov-Smirnoff test: $p - value > 0.05$, no statistically significant difference between observed and imputed values

# Appendix B Robustness

## Appendix B.1 Spatial PSVAR

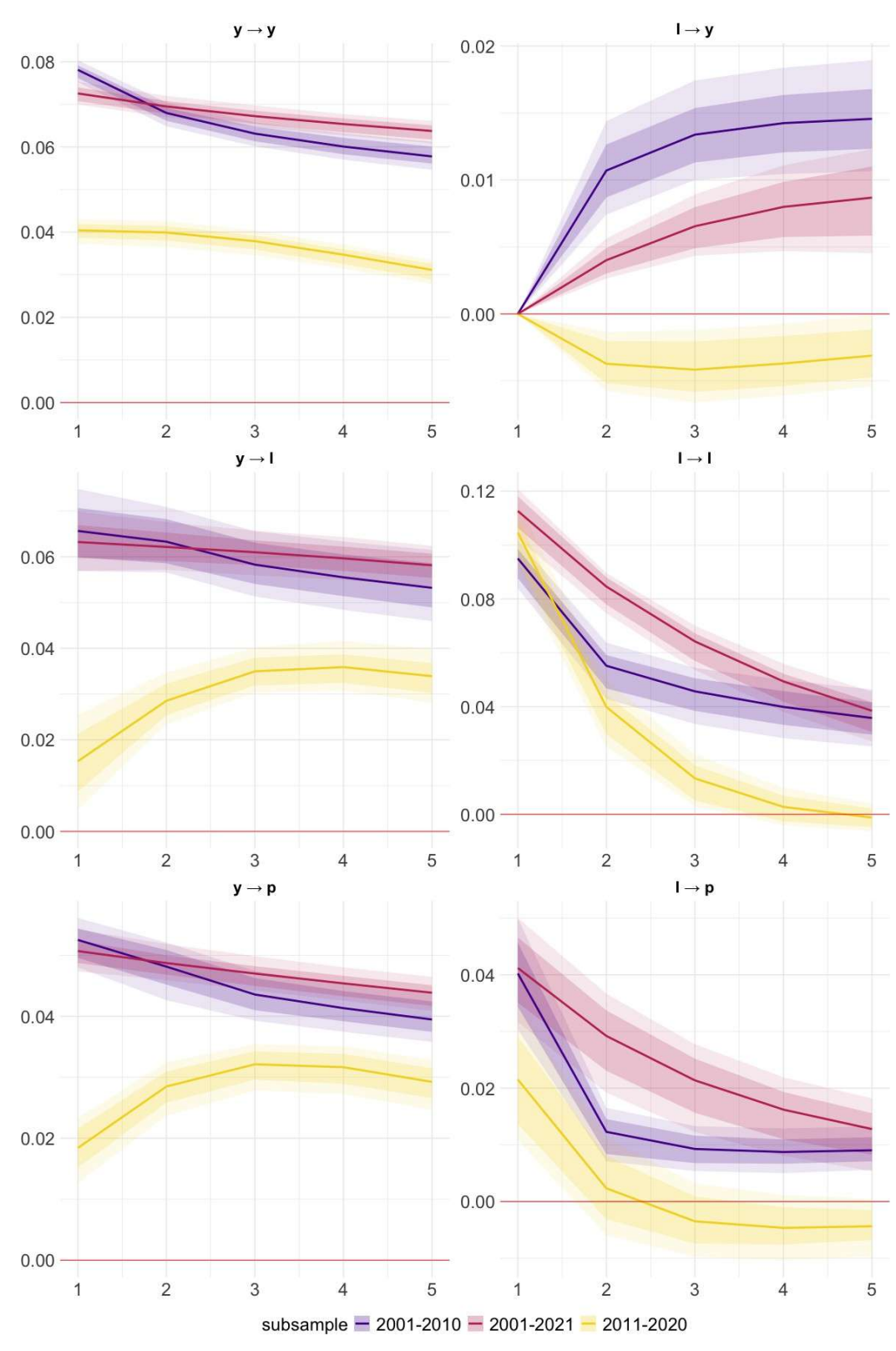

Figure 8: **IRFs Spatial PSVAR, Model 1 by sub-period (2001-2021; 2001-2010; 2011-2020):** Figures display IRFs of output ($y$), investment per worker ($I$), and labour productivity ($p$) to output ($y$) and investment per worker ($I$) shocks. Years on x-axis. Shaded areas denote 90% and 68% confidence bands calculated through m.b. bootstrapping (1000 runs).

| | 1 | 2 | 3 | 4 | 5 | Avg |
|---|---|---|---|---|---|---|
| | | | 2001-2021 | | | |
| $y$ | **0.70** | **0.70** | **0.70** | **0.70** | **0.70** | 0.70 |
| $I$ | **0.37** | **0.36** | **0.35** | **0.35** | **0.35** | 0.35 |
| | | | 2001-2010 | | | |
| $y$ | **0.67** | **0.69** | **0.69** | **0.69** | **0.69** | 0.69 |
| $I$ | **0.42** | **0.35** | **0.32** | **0.30** | **0.29** | 0.34 |
| | | | 2011-2020 | | | |
| $y$ | **0.46** | **0.58** | **0.67** | **0.72** | **0.76** | 0.64 |
| $I$ | **0.21** | 0.17 | 0.13 | 0.10 | 0.07 | 0.13 |

Table 10: **Cumulative effects estimated for Models 1 using Spatial PSVAR by sub-period (2001-2021; 2001-2010; 2011-2020).** Statistically significant values in bold (90% c.i.).

## Appendix B.2 Excluding the Covid-19 Pandemic (2001-2019)

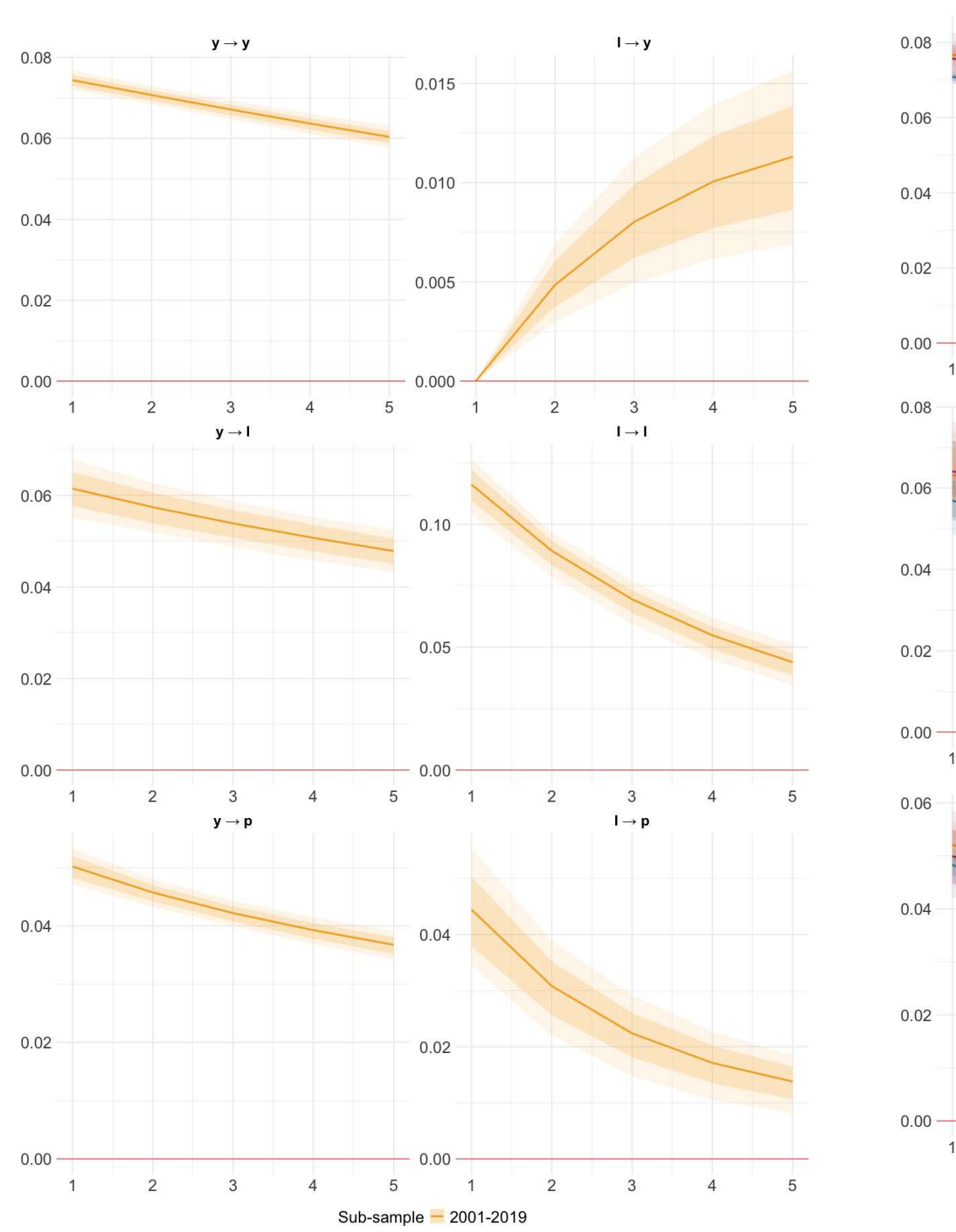

Figure 9: **IRFs, Model 1 excluding the Covid-19 Pandemic (2001-2019):** Figures display IRFs of output (*y*), investment per worker (*I*) and labour productivity (*p*) to output (*y*) and investment per worker (*I*) shocks. Years on x-axis. Shaded areas denote 90% and 68% confidence bands calculated through m.b. bootstrapping (1000 runs).

Figure 10: **IRFs, Model 1 by macro-region (East; Central; West) excluding the Covid-19 Pandemic (2001-2019):** Figures display IRFs of output (*y*), investment per worker (*I*) and labour productivity (*p*) to output (*y*) and investment per worker (*I*) shocks. Years on x-axis. Shaded areas denote 90% and 68% confidence bands calculated through m.b. bootstrapping (1000 runs).

| | 1 | 2 | 3 | 4 | 5 | Avg |
|---|---|---|---|---|---|---|
| | | | 2001-2019 | | | |
| *y* | **0.67** | **0.66** | **0.65** | **0.64** | **0.64** | 0.65 |
| *I* | **0.38** | **0.37** | **0.36** | **0.35** | **0.34** | 0.36 |

Table 11: **Cumulative effects, Model 1 excluding the Covid-19 Pandemic (2001-2019):** Statistically significant values in bold (90% c.i.). The average effect is estimated across 5 years.

| | | 1 | 2 | 3 | 4 | 5 | Avg |
|---|---|---|---|---|---|---|---|
| | | | | 2001-2019 | | | |
| **East** | *y* | **0.68** | **0.66** | **0.64** | **0.63** | **0.62** | 0.65 |
| | *I* | **0.61** | **0.56** | **0.53** | **0.50** | **0.48** | 0.53 |
| **Central** | *y* | **0.66** | **0.65** | **0.65** | **0.65** | **0.64** | 0.65 |
| | *I* | **0.32** | **0.31** | **0.31** | **0.31** | **0.31** | 0.31 |
| **West** | *y* | **0.68** | **0.67** | **0.66** | **0.65** | **0.64** | 0.66 |
| | *I* | **0.26** | **0.25** | **0.25** | **0.25** | **0.25** | 0.25 |

Table 12: **Cumulative effects estimated for Models 1 by macro-region (East; Central; West) excluding the Covid-19 Pandemic (2001-2019).** Statistically significant values in bold (90% c.i.). The average effect is estimated across 5 years.

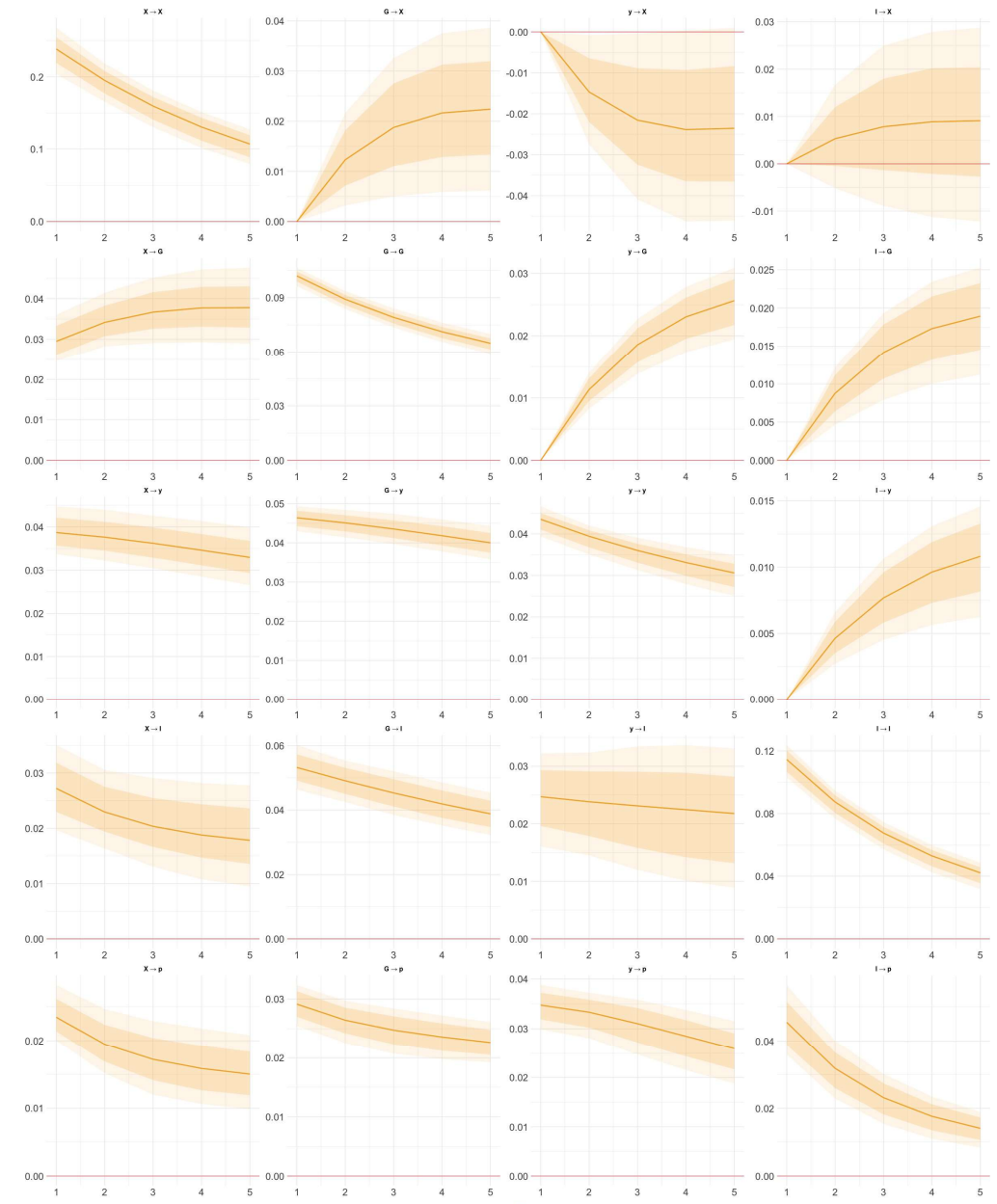

Figure 11: **IRFs, Model 2 excluding the Covid-19 Pandemic (2001-2019):** Figures display IRFs of export ($X$), government expenditure ($G$), output ($y$), investment per worker ($I$) and labour productivity ($p$) to export ($X$), government expenditure ($G$), output ($y$), and investment per worker ($I$) shocks. Years on x-axis. Shaded areas denote 90% and 68% confidence bands calculated through m.b. bootstrapping (1000 runs).

| | **1** | **2** | **3** | **4** | **5** | **Avg** |
|---|---|---|---|---|---|---|
| | | | **2001-2019** | | | |
| $X$ | **0.10** | **0.10** | **0.10** | **0.11** | **0.11** | 0.10 |
| $G$ | **0.29** | **0.29** | **0.30** | **0.30** | **0.31** | 0.30 |
| $y$ | **0.80** | **0.82** | **0.83** | **0.84** | **0.84** | 0.83 |
| $I$ | **0.40** | **0.38** | **0.37** | **0.37** | **0.36** | 0.38 |

Table 13: **Cumulative effects, Model 2 excluding the Covid-19 Pandemic (2001-2019):** Statistically significant values in bold (90% c.i.). The average effect is estimated across 5 years.

| | **1** | **2** | **3** | **4** | **5** | **Avg** |
|---|---|---|---|---|---|---|
| | | | **2001-2019** | | | |
| $X$ | **0.71** | **0.77** | **0.83** | **0.89** | **0.95** | 0.83 |
| $G$ | **2.42** | **2.55** | **2.66** | **2.76** | **2.84** | 2.65 |

Table 14: **Cumulative multipliers of output to export ($X$) and government expenditure ($G$) shocks, Model 2 excluding the Covid-19 Pandemic (2001-2019):** Statistically significant values (90% confidence bands) are in bold. Cumulative multipliers are computed as the cumulative change in output over the cumulative change in the respective variable across horizons up to 5 years. The average multiplier represents the mean cumulative effect over the 5-year horizon.

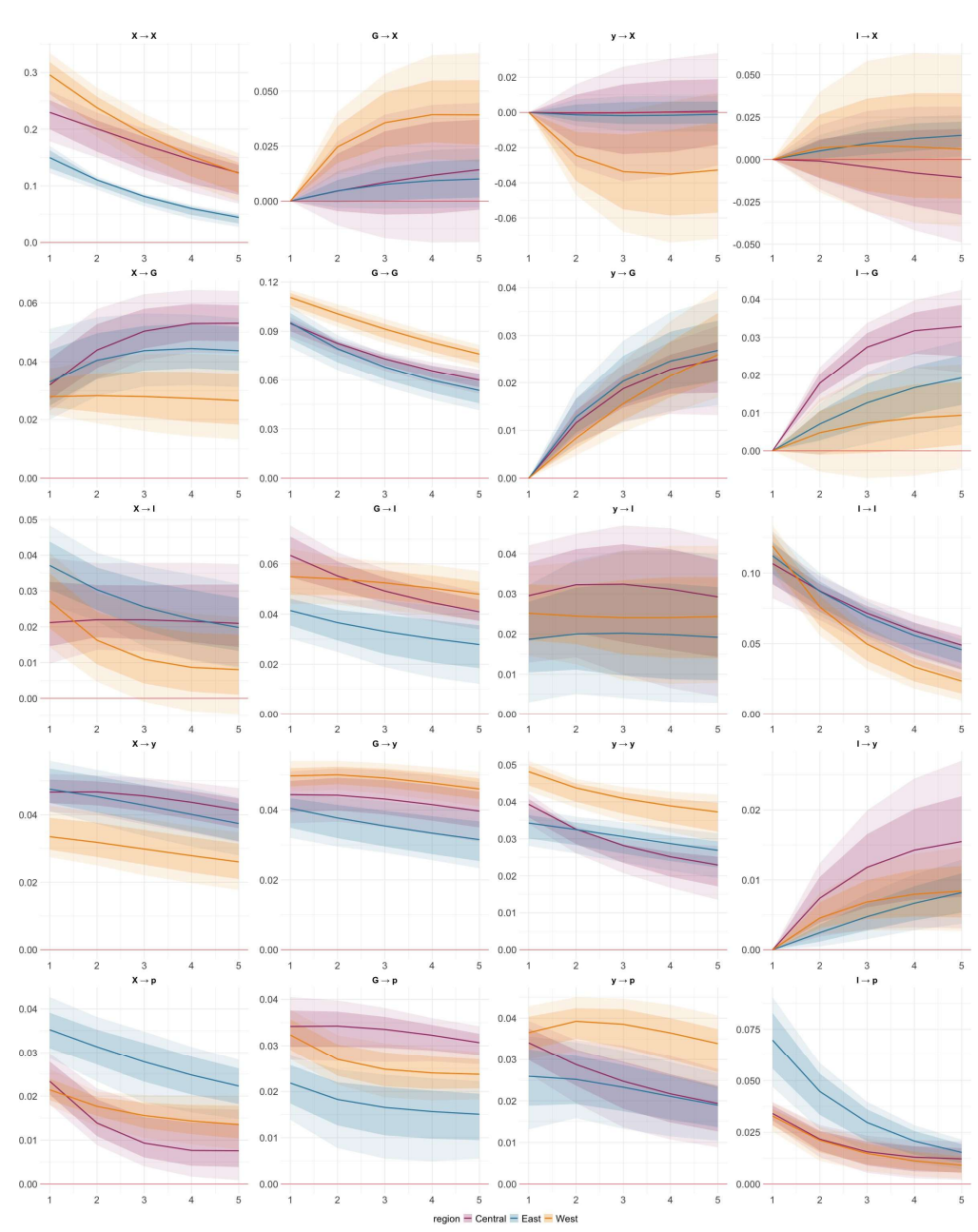

Figure 12: **IRFs, Model 2 by macro-region (East; Central; West) excluding the Covid-19 Pandemic (2001-2019):** Figures display IRFs of export ($X$), government expenditure ($G$), output ($y$), investment per worker ($I$) and labour productivity ($p$) to export ($X$), government expenditure ($G$), output ($y$), and investment per worker ($I$) shocks. Years on x-axis. Shaded areas denote 90% and 68% confidence bands calculated through m.b. bootstrapping (1000 runs).

| | | 1 | 2 | 3 | 4 | 5 | Avg |
|---|---|---|---|---|---|---|---|
| | | | | **2001-2019** | | | |
| East | $X$ | **0.24** | **0.26** | **0.28** | **0.30** | **0.32** | 0.28 |
| | $G$ | **0.23** | **0.23** | **0.23** | **0.24** | **0.25** | 0.24 |
| | $y$ | **0.76** | **0.77** | **0.76** | **0.76** | **0.75** | 0.76 |
| | $I$ | **0.62** | **0.57** | **0.54** | **0.51** | **0.49** | 0.55 |
| Central | $X$ | **0.10** | **0.09** | **0.08** | **0.07** | **0.07** | 0.08 |
| | $G$ | **0.36** | **0.39** | **0.41** | **0.42** | **0.44** | 0.40 |
| | $y$ | **0.87** | **0.87** | **0.87** | **0.87** | **0.87** | 0.87 |
| | $I$ | **0.32** | **0.29** | **0.27** | **0.26** | **0.26** | 0.28 |
| West | $X$ | **0.07** | **0.07** | **0.08** | **0.08** | **0.08** | 0.08 |
| | $G$ | **0.29** | **0.28** | **0.28** | **0.28** | **0.29** | 0.28 |
| | $y$ | **0.76** | **0.82** | **0.86** | **0.88** | **0.88** | 0.84 |
| | $I$ | **0.28** | **0.28** | **0.28** | **0.29** | **0.29** | 0.28 |

Table 15: **Cumulative effects, Model 2 by macro-region (East; Central; West) excluding the Covid-19 Pandemic (2001-2019):** Statistically significant values in bold (90% c.i.). The average effect is estimated across 5 years.

| | | 1 | 2 | 3 | 4 | 5 | Avg |
|---|---|---|---|---|---|---|---|
| | | | | **2001-2019** | | | |
| East | $X$ | **2.88** | **3.03** | **3.17** | **3.29** | **3.40** | 3.15 |
| | $G$ | **0.87** | **0.98** | **1.09** | **1.20** | **1.31** | 1.09 |
| Central | $X$ | **2.33** | **2.48** | **2.61** | **2.72** | **2.81** | 2.59 |
| | $G$ | **2.65** | **2.83** | **3.01** | **3.18** | **3.35** | 3.00 |
| West | $X$ | **1.63** | **1.71** | **1.78** | **1.84** | **1.90** | 1.77 |
| | $G$ | **1.50** | **1.62** | **1.74** | **1.86** | **1.97** | 1.74 |

Table 16: **Cumulative multipliers of output to export ($X$) and government expenditure ($G$) shocks, Model 2 by macro-region (East; Central; West) excluding the Covid-19 Pandemic (2001-2019):** Statistically significant values (90% confidence bands) are in bold. Cumulative multipliers are computed as the cumulative change in output over the cumulative change in the respective variable across horizons up to 5 years. The average multiplier represents the mean cumulative effect over the 5-year horizon.

# Appendix C  Complete Results

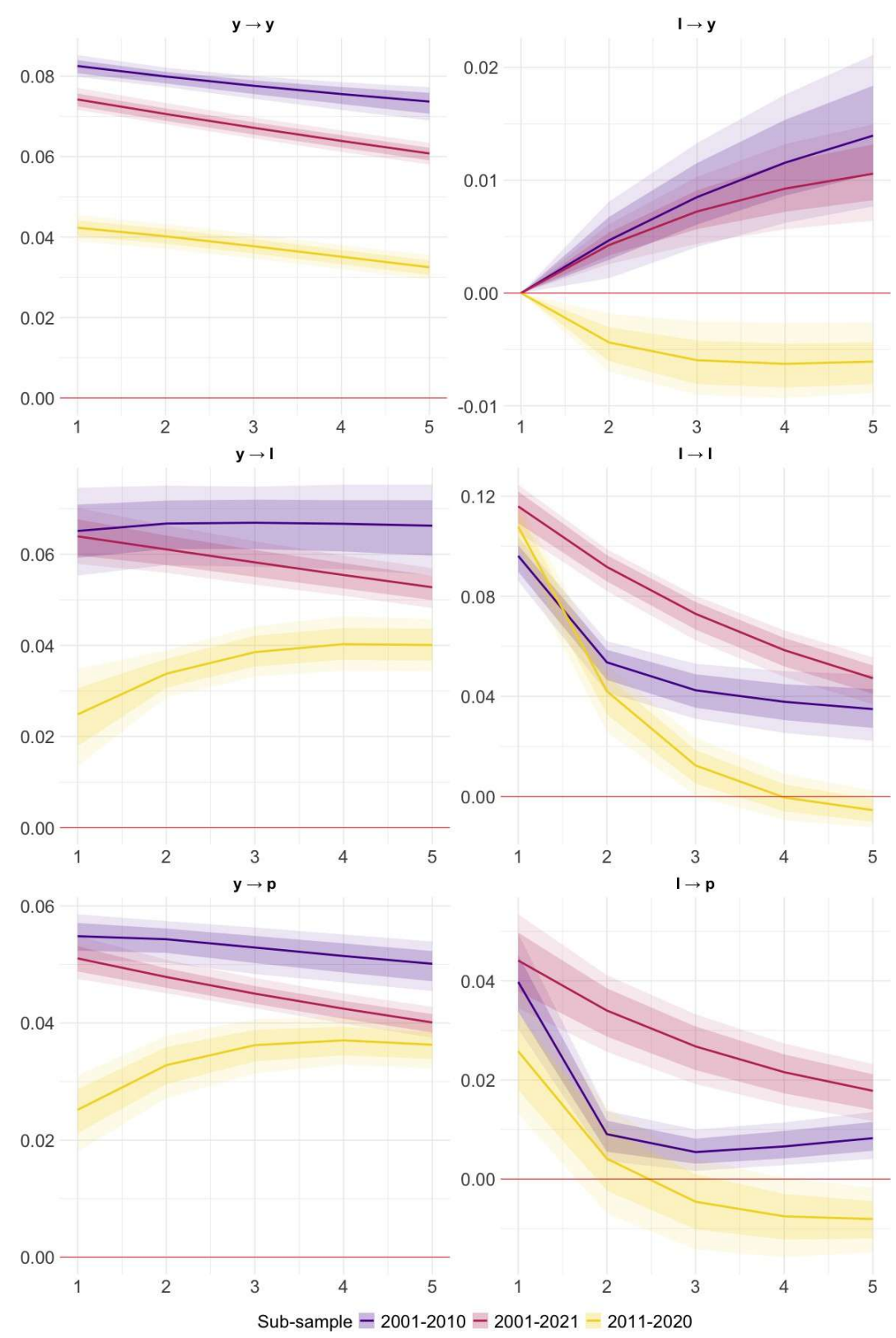

Figure 13: **IRFs, Model 1 by sub-period (2001-2021; 2001-2010; 2011-2020):** Figures display IRFs of output ($y$), investment per worker ($I$) and labour productivity ($p$) to output ($y$) and investment per worker ($I$) shocks. Years on x-axis. Shaded areas denote 90% and 68% confidence bands calculated through m.b. bootstrapping (1000 runs).

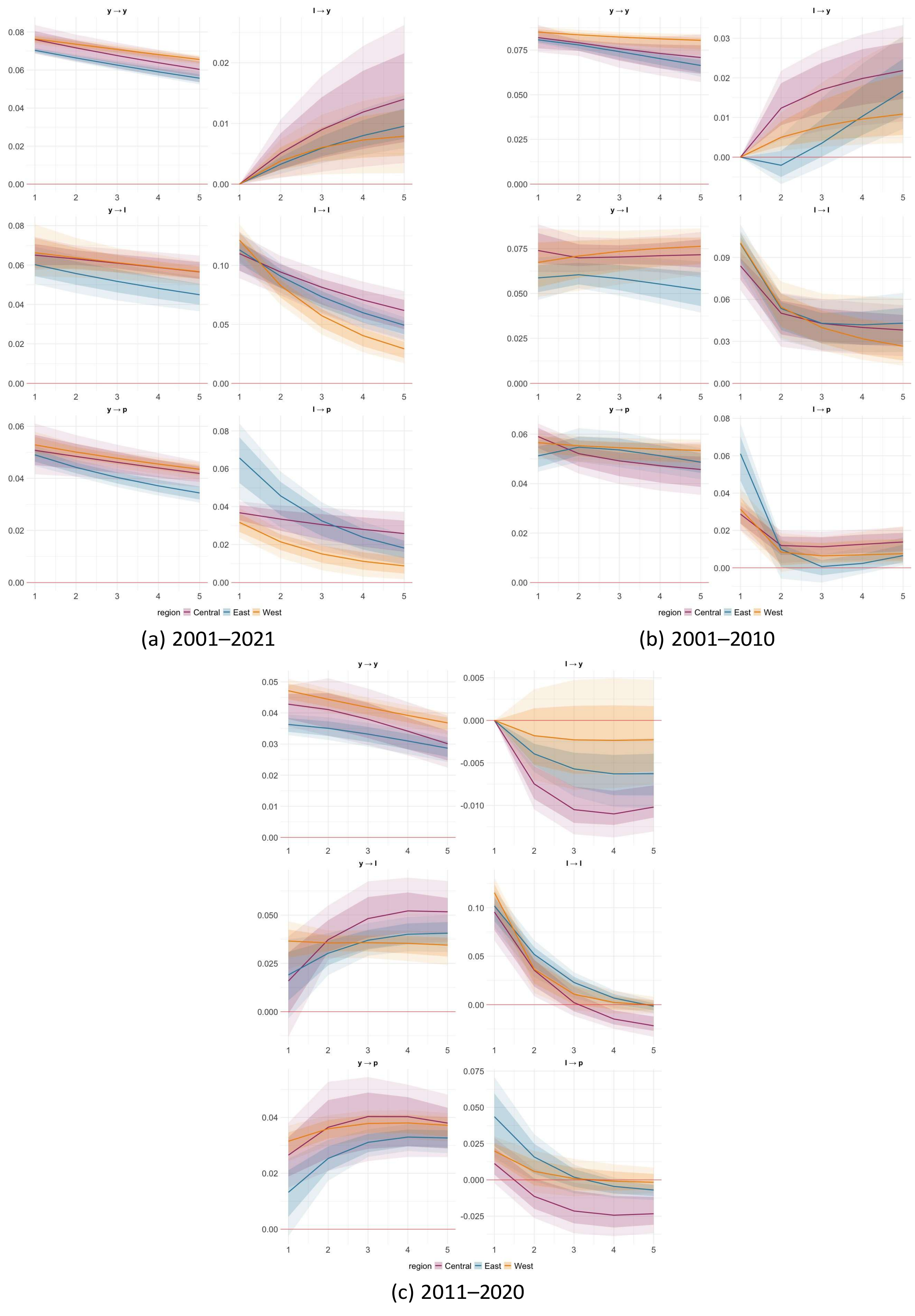

Figure 14: **IRFs, Model 1 by macro-region (East; Central; West) and sub-period (2001-2021; 2001-2010; 2011-2020):** Figures display IRFs of output (*y*), investment per worker (*I*) and labour productivity (*p*) to output (*y*) and investment per worker (*I*) shocks. Years on x-axis. Shaded areas denote 90% and 68% confidence bands calculated through m.b. bootstrapping (1000 runs).

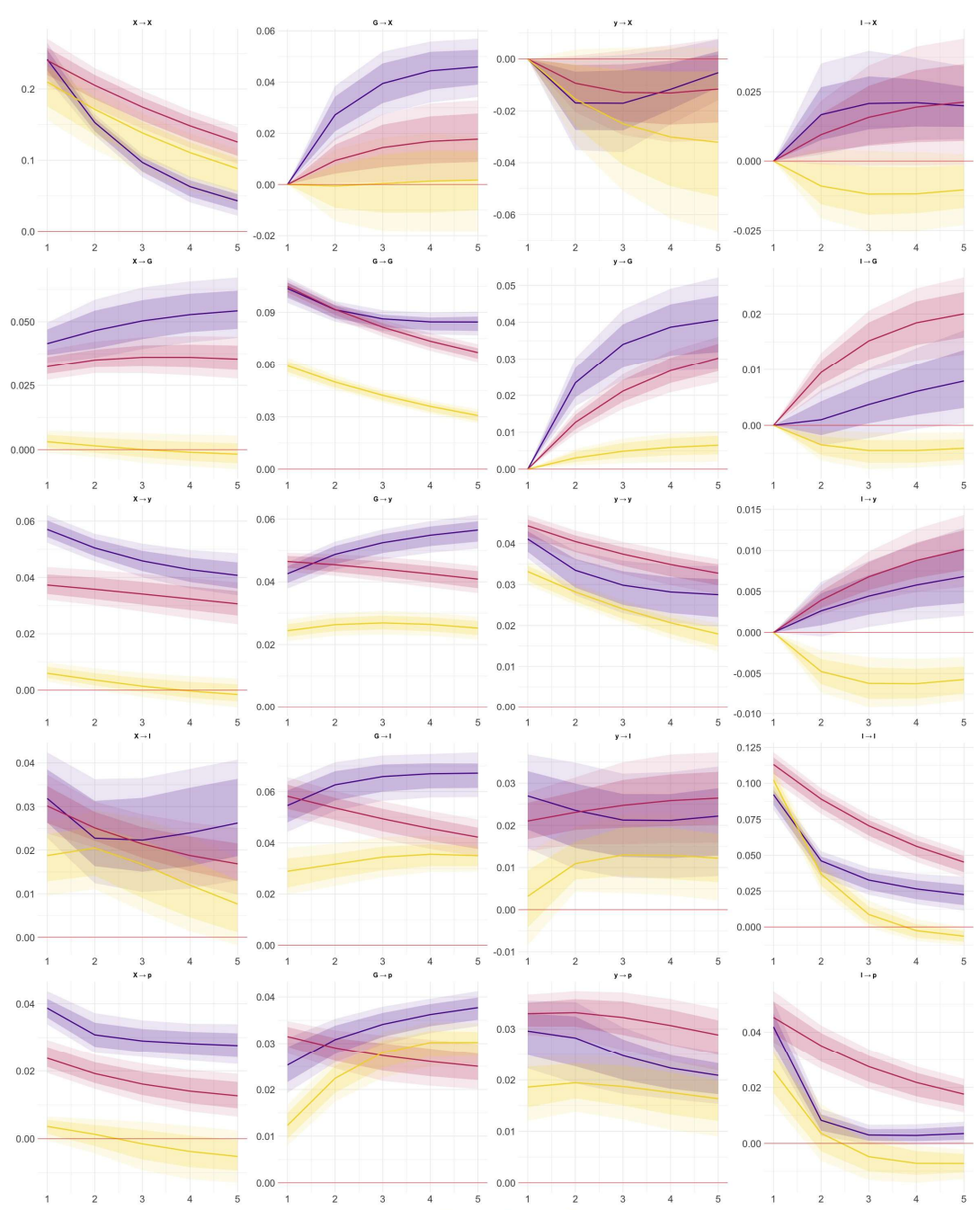

Figure 15: **IRFs, Model 2 by sub-period (2001-2021; 2001-2010; 2011-2020):** Figures display IRFs of export ($X$), government expenditure ($G$), output ($y$), investment per worker ($I$) and labour productivity ($p$) to export ($X$), government expenditure ($G$), output ($y$), and investment per worker ($I$) shocks. Years on x-axis. Shaded areas denote 90% and 68% confidence bands calculated through m.b. bootstrapping (1000 runs).

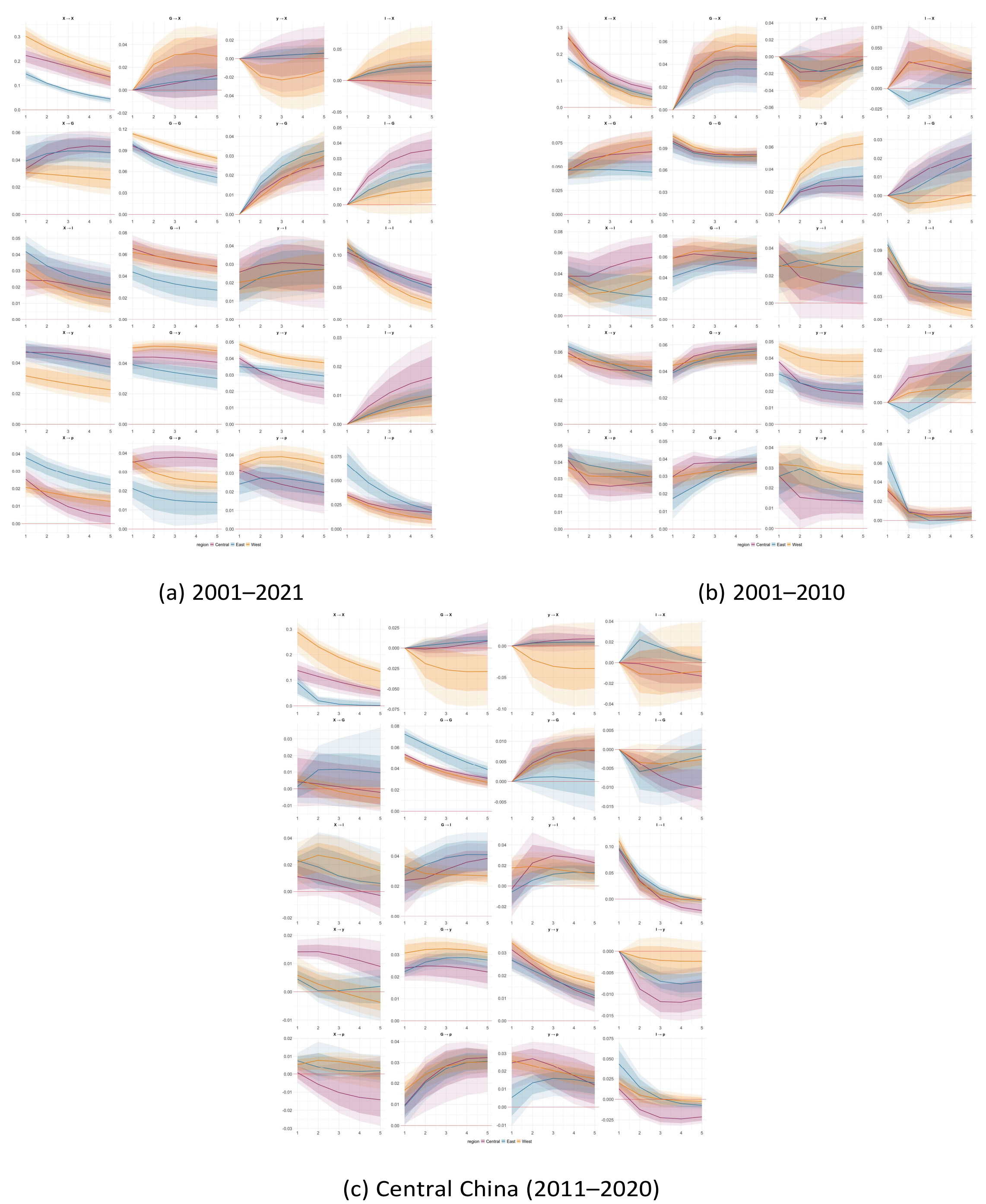

(a) 2001–2021

(b) 2001–2010

(c) Central China (2011–2020)

Figure 16: **IRFs, Model 2 by macro-region (East; Central; West) and sub-period (2001-2021; 2001-2010; 2011-2020):** Figures display IRFs of output to export ($X$) and government expenditure ($G$) shocks. Years on x-axis. Shaded areas denote 90% and 68% confidence bands calculated through m.b. bootstrapping (1000 runs).